\documentclass[letterpaper,11pt]{article}
\pdfoutput=1

\usepackage{xspace}
\usepackage{slashed}
\usepackage{array,multirow}
\usepackage{graphicx}
\usepackage{graphics}
\usepackage{epsfig}
\usepackage{amsfonts}
\usepackage{amsmath}
\usepackage{amssymb, bbold}
\usepackage{dsfont}
\usepackage{color}
\usepackage{xcolor}
\usepackage{cite}
\usepackage{pdflscape}
\usepackage{pifont}
\usepackage{pgfplots}
\usepackage{tikz} 
\usepackage{pgfplotstable}
\usepackage{enumitem}
\usepackage{cleveref}
\usepackage[normalem]{ulem}

\usepackage{bm}  %

\newcommand{\be}{\begin{equation}}
\newcommand{\ee}{\end{equation}}
\newcommand{\eps}{\epsilon}
\newcommand{\vp}{\varphi}
\renewcommand\({\left(}
\renewcommand\){\right)}
\renewcommand\[{\left[}
\renewcommand\]{\right]}

\newcommand{\dd}{{\rm d}}
\newcommand{\e}{{\rm e}}
\newcommand{\tr}{{\rm tr}}

\def\Tr{{\rm Tr}}
\def\nn{\nonumber}

\newcommand{\CPV}{\textsc{cp}\!\!\!\!\!\!\raisebox{0pt}{\small$\diagup$}}
\renewcommand{\Re}{{\rm Re}}
\renewcommand{\Im}{{\rm Im}}
\def\E{{\cal E}}

\usepackage{tikz}
\usetikzlibrary{arrows,shapes}
\usetikzlibrary{trees}
\usetikzlibrary{matrix,arrows} 				
\usetikzlibrary{positioning}				
\usetikzlibrary{calc,through}				
\usetikzlibrary{decorations.pathreplacing}  
\usepackage{pgffor}							

\usetikzlibrary{decorations.pathmorphing}	
\usetikzlibrary{decorations.markings}
\tikzset{
    vector/.style={decorate, decoration={snake}, draw},
	provector/.style={decorate, decoration={snake,amplitude=2.5pt}, draw},
	antivector/.style={decorate, decoration={snake,amplitude=-2.5pt}, draw},
    fermion/.style={draw=black, postaction={decorate},
        decoration={markings,mark=at position .55 with {\arrow[draw=black]{>}}}},
    fermionbar/.style={draw=black, postaction={decorate},
        decoration={markings,mark=at position .55 with {\arrow[draw=black]{<}}}},
    fermionnoarrow/.style={draw=black},
    gluon/.style={decorate, draw=black,
        decoration={coil,amplitude=4pt, segment length=5pt}},
    scalar/.style={dashed,draw=black, postaction={decorate},
        decoration={markings,mark=at position .55 with {\arrow[draw=black]{>}}}},
    scalarbar/.style={dashed,draw=black, postaction={decorate},
        decoration={markings,mark=at position .55 with {\arrow[draw=black]{<}}}},
    scalarnoarrow/.style={dashed,draw=black},
    electron/.style={draw=black, postaction={decorate},
        decoration={markings,mark=at position .55 with {\arrow[draw=black]{>}}}},
	bigvector/.style={decorate, decoration={snake,amplitude=4pt}, draw},
}

\begin{document}

\begin{titlepage}

\begin{flushright}
Nikhef 2019-047

DESY 19-182
\end{flushright}

\vspace{2.0cm}

\begin{center}
{\LARGE  \bf 
 Source terms for electroweak baryogenesis in the vev-insertion approximation beyond leading order
}

\vspace{2.4cm}

{\large \bf Marieke Postma$^a$ and Jorinde van de Vis$^{a,b}$ 
}
\vspace{0.5cm}

\vspace{0.25cm}

{\large 
$^a$ 
{\it Nikhef, Theory Group, Science Park 105, 1098 XG, Amsterdam, The Netherlands}}

\vspace{0.25cm}
{\large 
$^b$ 
{\it DESY, Notkestra{\ss}e 85, D-22607, Hamburg, Germany}
}

\end{center}

\vspace{1.5cm}

\begin{abstract}
  In electroweak baryogenesis the baryon asymmetry of the universe is created during the electroweak phase transition. The quantum transport equations governing the dynamics of the plasma particles can be derived in the vev-insertion approximation, which treats the vev-dependent part of the particle masses as a perturbation.  We calculate the next-to-leading order (NLO) contribution to the CP-violating source term and CP-conserving relaxation rate, corresponding to Feynman diagrams for the self-energies with four mass insertions.  We consider both a pair of Weyl fermions and a pair of complex scalars, that scatter off the bubble wall. We find: (i) The NLO correction becomes large for ${\cal O}(1)$ couplings. If only the Standard Model (SM) Higgs obtains a vev during the phase transition, this implies the vev-insertion approximation breaks down for top quarks. (ii) The resonant enhancement of the source term and relaxation rate, that exists at leading order in the limit of degenerate thermal masses for the fermions/scalars, persists at NLO.
\end{abstract}

\vfill
\end{titlepage}

\tableofcontents
\newpage

\section{Introduction}\label{sec:intro}

In electroweak baryogenesis (EWB) the baryon asymmetry of the universe is created during the electroweak phase transition.  The scenario requires new physics at the electroweak scale, in particular an extension of the standard model (SM) scalar sector to obtain a strong first-order phase transition, and new sources of CP violation beyond those present in the CKM matrix. A major motivation to study EWB in detail is that it can be probed by experiment,  for instance by collider searches for new scalars \cite{Arhrib:2013oia, Chen:2013rba, Chang:2017ynj, CMS:2016tgd}, precision Higgs studies \cite{Englert:2014uua, Brivio:2017vri}, and CP-odd collider observables \cite{Han:2009ra, Boudjema:2015nda, Ellis:2015dha, Askew:2015mda, Demartin:2015uha}. Particularly tight constraints on new sources of CP violation come from measurements of the electric dipole moment \cite{Chupp:2017rkp, Balazs:2016yvi,Chien:2015xha,Cirigliano:2016nyn, deVries:2017ncy,Andreev:2018ayy}. In addition, gravitational waves produced during the phase transition may be measured by LISA or other future gravitational wave observatories \cite{Audley:2017drz, Caprini:2015zlo}.

To draw definite conclusions on the viability of EWB scenarios, accurate theoretical predictions are needed, which are currently somewhat lacking.  Although no comprehensive comparison of different theoretical approaches exists, it has been shown for specific models that predictions may vary by more than an order of magnitude \cite{Konstandin:2013caa}.  The computation is hampered by the non-equilibrium, non-perturbative, and finite-temperature aspects of the process.  The starting point for a quantum treatment of EWB is the closed-time-path formalism at finite temperature.  The dynamics of the phase space densities of plasma quanta are described by the Kadanoff-Baym equations \cite{KaBa, Prokopec:2003pj, Calzetta:2008iqa}. Unfortunately, these equations are complicated, and a series of approximations -- which are not always controlled -- are needed to make progress.

We will focus on the vev-insertion approximation (VIA) method \cite{Huet:1995mm, Riotto:1998zb, Lee:2004we}. To derive a set of transport equations for the plasma particles that are simple enough to solve (numerically), the following main approximations are made: (i) The bubble dynamics is treated as slow compared to the typical time-scale of the plasma excitations. (ii) Quantum coherence effects are neglected, which allows to rewrite the  Kadanoff-Baym equations in terms of number densities rather than phase-space densities.  (iii) It is assumed that Fick's first law can be used to incorporate diffusion. And finally, (iv) the effective mass of the relevant plasma particles, which is spacetime-dependent in the bubble wall background, is treated as a perturbation.

The VIA method derives its name from the last approximation, as it corresponds to an expansion in the number of vev-insertions, that is, insertions of the two-point coupling, in the Feynman diagrams for the particles scattering in the bubble background. In this paper, we will determine the validity of this expansion, by calculating the next-to-leading order (NLO) correction to the CP-odd and -even rates for fermion/scalar interactions with the bubble wall. We will consider both a left- and right-handed fermion pair with a Yukawa-type interaction with the Higgs field, and a pair of complex scalars -- also denoted by left and right -- with a left-right-mixing coupling to the Higgs.

The transport equation for the right-handed field is of the form
\be
\partial_\mu j_R^\mu=  S^{\CPV}  -\Gamma^+( \mu_R +\mu_L)  -\Gamma^-( \mu_R -\mu_L)  -\Gamma_H( \mu_R -\mu_L- c\; \mu_H) + ...\, .
\label{transport}
\ee
Similar equations exist for the left-handed field and for the Higgs boson.  Here $j_R^\mu$ is the number current for the right-handed particles (the zeroth component corresponds to the
number density of particles minus antiparticles), $\mu_i$ the chemical potential of particle $i$, and $c$ depends on the coupling to the Higgs field.  The CP-violating term  $S^{\CPV}$ sources the creation of a non-zero number density of right-handed particles, and it originates from interactions with the bubble wall.
All other rates on the right-hand side of the equation conserve CP, and they give rise to washout of the number density.  The relaxation rates $\Gamma^\pm$ encode interactions with the bubble wall, $\Gamma_H$ interactions with the Higgs quanta, and the ellipses stand for all possible other interactions.  A net chiral density in front of the bubble wall is transformed into a net baryon asymmetry by weak sphaleron transitions.

We will calculate the CP-even rates  $\Gamma^\pm$ and the CP-odd source term $S^{\CPV}$ to NLO in the vev-insertion expansion. They arise from the scatterings of the fermions/scalars with the bubble wall, mediated by the vev-dependent two-point coupling, $f_f(\vp_b)$ and $f_s(\vp_b)$ for fermions and scalars respectively, with $\vp_b$ the bubble wall Higgs background. Let's focus on the fermions. Based on naive dimensional analysis one would expect the NLO correction to be suppressed by a factor $|f_f|^2/T_N^2$,  with $T_N$ the temperature at bubble nucleation.  However, it is not obvious that this estimate is accurate as there are other scales in the problem, such as the thermal decay widths $\Gamma_T$ of the particles.  In fact, at leading-order in VIA the dominant rate and source term are both  resonantly enhanced in the limit that the field-independent (thermal) masses for the left- and righthanded particles are degenerate, and then  scale as $S^{\CPV}, \Gamma^-\propto 1/\Gamma_T$. Moreover, as we will show, since the correction can be order one for the top quark, it is important to check for possible numerical factors in the expansion parameter. It should be noted that the vev-insertion expansion is not a loop expansion and thus there are no factors of $(4\pi)$ appearing.

The NLO expressions generically have a complicated form, but simplify enormously in the limit of degenerate masses for the left- and right-handed particles, which is a good approximation for the top quark.  We then find that the ratio of the NLO to LO contribution for fermions is ($\Gamma^+$ vanishes in the mass degenerate limit):
\be
\frac{|\Gamma^-|_\textsc{nlo}}{|\Gamma^-|_\textsc{lo}}
=\frac{|S^{\CPV}|_\textsc{nlo}}{|S^{\CPV}|_\textsc{lo}}
\sim\frac{|f_f|^2m_T^2}{8 T_N^2 \Gamma_T^2} \sim \frac{|f_f|^2}{8 \alpha T_N^2} \, ,
\label{Result}
\ee
with $m_T$ the thermal mass, and $\alpha$ the QCD (electroweak) coupling for fermions with strong (electroweak) interactions.  For scalars the enhancement factor also depends on the thermal width, and is given explicitly in \cref{G_estimate,S_estimate}.  For quarks \cref{Result} is of the order of the naive estimate $|f_f|^2/T_N^2$. The couplings $f_i$ are model dependent, and the validity of the vev-insertion approximation should be checked per model.  In set-ups where only the SM Higgs is non-zero along the bubble wall $|f_f| \approx y_f \vp_b$ with $y_f$ the SM Yukawa coupling, and thus VIA breaks down for a pair of top quarks, but works well for all other SM fermions.

We further find that in the degenerate mass limit the NLO correction is just a multiplicative factor.  The resonant condition  of the leading order source term and relaxation rate is not shifted or otherwise affected, and is a robust result.

This paper is organized as follows. In the next section, we briefly summarize the relevant formalism. We first introduce the fermionic and scalar set-up and define the two-point coupling $f_f$ and $f_s$  in \cref{s:model}. \Cref{s:formalism} recaps the Feynman rules and propagator relations relevant for the calculation of the source terms in the CTP formalism, and \cref{s:transport} gives the transport equations in the vev-insertion approximation. To calculate the source terms, multiple contour integrals need to be performed. To set the notation, we give the master integrals in \cref{s:contour}. In \cref{s:LO} we review the leading order calculation for both the CP-violating and -conserving source terms. In \cref{s:NLO} the calculation is extended to the next-to-leading order contribution. We will provide all the calculational details. Readers only interested in the final results can go from \cref{s:transport} straight to \cref{s:results} where we summarize the outcome of the NLO calculation, and discuss the implications. 


\section{Set-up, formalism and notation }
\label{s:set-up}

In this section we introduce the set-up, and recap the formalism to calculate the source $S^{\CPV}$ and wash-out rates  $\Gamma^\pm$ in the transport equation (\ref{transport}) from Feynman diagrams. The leading order and next-to-leading order calculation is then presented in the next sections.

\subsection{Model}
\label{s:model}

We consider a two-flavor\footnote{In this work ``flavor'' means $L,R$ (as in Ref. \cite{Konstandin:2004gy}).} system consisting of either a pair of chiral fermions ($\psi_L,\,\psi_R$) or a pair of complex scalars ($\phi_L,\,\phi_R$), and we are interested in their dynamics during the electroweak phase transition. In the high temperature expansion the thermal masses of the particles are field independent, and they are included in the free Lagrangian, and thus appear in the propagators defined in \cref{s:formalism}. Scalars can in addition have a flavor diagonal constant mass term; this possibility can be included by substituting everywhere $m_T^2 \to m_T^2 + M^2$ with $M$ the bare mass.
  CP-violation resides in left-right mixing couplings to the Higgs field, which in the bubble wall background generates a field-dependent mass for the fermions/scalars.  This mass is treated as a perturbation in the vev-insertion approximation (VIA), and included in the interaction Lagrangian. The Higgs vev can be parametrized as $\langle H^\dagger H \rangle = \frac12\vp_b^2(x^\mu)$, with $\vp_b(x^\mu)$ the space-time dependent bounce solution (in multi-Higgs models, we are interested in the linear combination of Higgs vevs that enters the bounce solution for tunneling).  We can then write the two-point interaction for the fermions and scalars as
\begin{align}
  {\cal L}^{\rm int}_f \supset - \frac{f_f(\vp_b)}{\sqrt{2}} \bar\psi_L\psi_R
  + {\rm h.c.}  , \qquad
  {\cal L}_s^{\rm int} \supset 
  - f_s(\vp_b)  \phi_L^\dagger \phi_R + {\rm h.c.}  ,
  \label{define_f}
\end{align}
with $f_i$ and $i=f,s$ the CP-violating field-dependent mass term. It describes the scattering of the plasma particles off the bubble wall. The interaction violates CP, and particles and antiparticles scatter differently,  provided $\Im(f_i^*\dot f_i) \neq 0$. In explicit models, this can be achieved if two (or more) different interactions with complex couplings contribute to $f_i$, as then the relative phase cannot be rotated away in the bubble background.  For example, CP-violation (CPV) can arise in a two-Higgs doublet model from interference between couplings  to the two Higgs fields  \cite{Bochkarev:1990fx,Turok:1991uc,Davies:1994id ,Cline:1996mga,Cline:2011mm,Dorsch:2016nrg, Andersen:2017ika,Gorda:2018hvi}. Alternatively, in an  effective field theory approach it can come from interference between SM Yukawa and dimension-six effective interactions \cite{Zhang:1994fb, Bodeker:2004ws, Balazs:2016yvi, deVries:2017ncy}. When we discuss the implications of our work in \cref{s:results} we will focus on the case where ($\psi_L,\,\psi_R$)  are the chiralities of a single SM fermion, e.g.  the  left- and right-handed top quark, but the results can be straightforwardly generalized to set-ups where the CP-violating interaction is between particles from different families \cite{Guo:2016ixx, Chiang:2016vgf}. For the scalar set-up the coupling $f_s$ can both be linear in the Higgs field (plus possible higher dimensional terms), as happens in supersymmetric models \cite{Lee:2004we}, and quadratic \cite{Cirigliano:2009yt}.

For our considerations, the origin of the mass term is not important, and we work with the phenomenological parametrization in \cref{define_f}.

\subsection{CTP formalism and Wightman functions}
\label{s:formalism}

We use the formalism of \cite{Lee:2004we}, and our metric is $(+,-,-,-,)$.

The evolution of the plasma quanta during the electroweak phase transition is described using the finite-temperature, non-equilibrium Closed Time Path (CTP) or Schwinger-Keldysh formalism \cite{Schwinger:1960qe, Keldysh:1964ud, Chou:1984es}. For an extensive introduction to the CTP formalism, see e.g. \cite{Calzetta:2008iqa}.  

All integrals and derivatives are performed along a closed path, that can be split into a plus-branch from the initial time (which for initial thermal equilibrium can be taken to minus infinity) to some finite time $t$, and a minus-branch going backwards. 
There are then four Green functions, depending on the branch that the time arguments of the fields are located on.
The interactions connect fields on the same branch; the Feynman rule is that every mass insertion gives a factor $-if_s^j$ for the scalar theory, with $j=\pm$ denoting the branch, and a factor $-if_f^j/\sqrt{2}$ for the fermionic theory. Since the minus branch runs backward in time the coupling picks up an additional sign
\be
f_i^+ = - f_i^- \equiv f_i.
\label{sign_f}
\ee

\paragraph{Scalars}  For scalars the Wightman functions are defined as
\be
G^{-+}(x,y) = G^>(x,y) = \langle \phi(x) \phi^\dagger(y) \rangle, \qquad
G^{+-}(x,y) = G^<(x,y) = \langle \phi^\dagger(y)\phi(x) \rangle \, ,
\ee
where we have  suppressed the labels $L,R$ on the fields and Wightman functions.  Since the thermal masses are flavor diagonal, and the off-diagonal mass term is treated as a perturbation, all Green functions are diagonal in flavor space.
The time- and anti-time-ordered propagators  are
\begin{align}
G^{++}(x,y) &=G^{t}(x,y) = \Theta(x^0-y^0) G^{-+}(x,y)+\Theta(y^0-x^0) G^{+-}(x,y) ,\nn \\
G^{--}(x,y) &=G^{\bar t}(x,y) = \Theta(x^0-y^0) G^{+-}(x,y)+\Theta(y^0-x^0) G^{-+}(x,y) ,
\label{Gplus}
\end{align}
with $\Theta(x)$ the Heaviside step function. Under complex conjugation the Green functions transform as
\be
(G_{xy}^{-+})^\dagger = G^{-+}_{yx}\,, \qquad
  (G_{xy}^{+-})^\dagger = G^{+-}_{yx}\,, \qquad
(G_{xy}^{++})^\dagger = G^{--}_{yx}\,, \qquad
(G_{xy}^{--})^\dagger = G^{++}_{yx} \, ,
\label{Gdagger}
\ee
where we introduced the notation $G_{xy} = G(x,y)$.

Explicitly, the Wightman functions are given by
\begin{align}
\label{Gk}
G^\lambda(x,y)=\int_k \, e^{-i{k}\cdot(x-y)}
g_s^\lambda({k}_0,\mu)\rho(k_0,\vec{k})
\,,
\end{align}
with $\lambda=>,<$ and $\int_k  = \int \dd^4 k/(2\pi)^4$. We will also use the notation $\int_{\vec k}  = \int \dd^3 \vec k/(2\pi)^3$ and $\int_{k^0} =\int \dd k^0/(2\pi)$. The spectral density is
\begin{align}
  \rho({k}_0, \vec{k})       
 &=  {i\over 2\omega_k}\[
{1\over
 {k}_0-\E^*}-{1\over {k}_0+\E}
 -{1\over {k}_0-\E}+
{1\over {k}_0+\E^*}\],
\label{rho}
\end{align}
with
\be
\E = \omega_k + i \Gamma_T, \quad
\E^* = \omega_k - i \Gamma_T.
\label{calE}
\ee
We include the (leading order) thermal corrections in the propagators and take $\omega_k^2 = {\vec k}^2 +m_T^2(T)$ with $m_T$ and $\Gamma_T$ the thermal mass and width respectively.  We neglect deviations from thermal equilibrium for the thermal distribution functions, and expand in small chemical potential:
\begin{align}
  g_s^>(k_0,\mu) &=1+n_s(k_0-\mu)= 1 +n_s(k_0) -\frac{\mu h_s(k_0)}{T}  + {\cal O}(\mu^2),
                     \nn \\
  g_s^<(k_0,\mu)&= n_s(k_0-\mu)=n_s(k_0) -\frac{\mu h_s(k_0)}{T} + {\cal O}(\mu^2),
                    \label{gexpB}
\end{align}
with the Bose-Einstein distribution and its derivative given by
\be
n_s(x) = \frac{1}{(\e^{x/T}-1)}, \qquad
h_s(x) =T n_s'(x) = - \frac{\e^{x/T}}{(\e^{x/T}-1)^2}.
\ee

\paragraph{Fermions} 

The fermionic Wightman functions are defined via
\be
S^{-+}(x,y) = S^>(x,y) = \langle \psi(x) \bar \psi(y) \rangle, \qquad
S^{+-}(x,y) = S^<(x,y) = \langle \bar \psi(y)\psi(x) \rangle \, ,
\ee
where we suppressed both spinor and flavor indices. 
The time-ordered propagators are defined by equivalent relations to the scalar ones \cref{Gplus}.  The hermiticity properties are
\be
(S_{xy}^{-+})^\dagger = \gamma^0 S^{-+}_{yx} \gamma^0\, ,\quad
  (S_{xy}^{+-})^\dagger =\gamma^0 S^{+-}_{yx} \gamma^0\, ,\quad
(S_{xy}^{++})^\dagger = \gamma^0 S^{--}_{yx} \gamma^0\, ,\quad
(S_{xy}^{--})^\dagger =\gamma^0 S^{++}_{yx}\gamma^0 .
\label{Sdagger}
\ee

The explicit form of the Wightman function is
\be
\label{Sk}
S^\lambda(x,y)=\int_k e^{-i{k}\cdot(x-y)}
g_f^\lambda({k}_0,\mu)\rho(k_0,\vec{k})\left(\slashed{k}+m_T\right)
\,,
\ee
with the spectral density again given by \cref{rho}, and $m_T$ the thermal mass.  For future reference we introduce the notation
\be
\label{SKtrace}
S^\lambda(x,y)\big|_{\Tr(m)=0}=\int_k e^{-i{k}\cdot(x-y)}
g_f^\lambda({k}_0,\mu_i)\rho(k_0,\vec{k})\left(\slashed{k}\right)
\,,
\ee
that is, the subscript $\Tr(m)=0$ indicates that the mass term in the last factor of \cref{Sk} is set to zero.
The thermal distribution functions for fermions are
\begin{align}
g_f^>(k_0,\mu_i)&=1-n_f(k_0-\mu_i) = 1 - n_f(k_0) +\frac{\mu h_f(k_0)}{T} + {\cal O}(\mu^2) \, ,\nn \\
  g_f^<(k_0,\mu_i)&= -n_f(k_0-\mu_i)=- n_f(k_0) +\frac{\mu h_f(k_0)}{T} + {\cal O}(\mu^2)\,,
                     \label{gexpF}
\end{align}
with the Fermi-Dirac distribution and its derivative given by
\be
n_f(x) = \frac1{(\e^{x/T}+1)},\qquad h_f(x) = T n_f'(x) =  -\frac{\e^{x/T}}{(\e^{x/T}+1)^2}.
\ee
%

\subsection{Transport equations}
\label{s:transport}

The transport equations can be derived directly from the Schwinger-Dyson equations \cite{Riotto:1998zb,Lee:2004we}, or equivalently, from the Kadanoff-Baym equations \cite{Dev:2014wsa}. The latter approach makes transparant that only the leading order terms in the derivative expansion are included, i.e., that it is assumed that the bubble wall dynamics is slow compared to the typical timescale of the plasma excitations which is set by the temperature. The result for scalars is
\begin{equation}
\partial_\mu j^\mu_s(x) =  
\int d^3 y\int_{-\infty}^{x_0} dy_0\
\Bigl[ \Pi^{-+}_{xy} G^{+-}_{yx}- \Pi^{+-}_{xy} G^{-+}_{yx}-G^{-+}_{xy}\Pi^{+-}_{yx}
+G^{+-}_{xy}\Pi^{-+}_{yx} \Bigr]\,,
\label{Jscalar}
\end{equation}
with $\Pi^\lambda_{xy} \equiv\Pi ^{\lambda}(x,y)$ the self-energy.  In the diffusion approximation $j^\mu=(n, -D \vec \nabla n)$ with $D$ the diffusion constant. Neglecting the bubble wall curvature, the transport equations reduce to ordinary differential equations for the number densities of plasma particles. 
The transport equation for a Dirac fermion current is
\begin{equation}
\partial_\mu j^\mu_f(x) =  
-\int d^3 y\int_{-\infty}^{x_0} dy_0\
{\rm Tr}\Bigl[ \Sigma^{-+}_{xy} S^{+-}_{yx}- \Sigma^{+-}_{xy} S^{-+}_{yx}-S^{-+}_{xy}\Sigma^{+-}_{yx}
+S^{+-}_{xy}\Sigma^{-+}_{yx} \Bigr]\,,
\label{Jfermion}
\end{equation}
with $\Sigma^\lambda_{xy}$ the self-energy.
Apart from the overall sign and the trace over spinor space, this has the same form as the scalar transport \cref{Jscalar}.

The transport equations can be simplified using the hermiticity properties.  The self-energies satisfy the same type of relations as the propagators in \cref{Gdagger,Sdagger} 
\be
(\Pi_{xy}^{-+})^\dagger = \Pi^{-+}_{yx} ,\qquad
  (\Pi_{xy}^{+-})^\dagger = \Pi^{+-}_{yx} ,\qquad
(\Sigma_{xy}^{-+})^\dagger = \gamma^0\Sigma^{-+}_{yx}\gamma^0 ,\qquad
  (\Sigma_{xy}^{+-})^\dagger = \gamma^0\Sigma^{+-}_{yx}\gamma^0.
\label{Pidagger}
\ee
We will verify this explicitly for the LO and NLO contributions in the next two sections.

We will always calculate the transport equation for the right-handed particle, the corresponding one for the left-handed particle then follows from $\partial_\mu j^\mu_{L,i}=-\partial_\mu j^\mu_{R,i}$ with $i=f,s$ for fermions respectively scalars.  Putting back the flavor indices, and using the hermiticity properties of the self-energy, this becomes
\begin{align}
  \partial_\mu j_{R,s}^\mu(x) &=\;\;\;  \int \dd^4 y \, \Theta_{xy}\, 2\Re [\Pi^{-+}_{R,xy} G^{+-}_{R,yx}- \Pi^{+-}_{R,xy} G^{-+}_{R,yx}  ] \;\;\;=S_{s}^\textsc{cp}+ S_{s}^{\CPV},\nn\\
  \partial_\mu j_{R,f}^\mu(x) &= - \int \dd^4 y \, \Theta_{xy}\, 2\Re \, \Tr [\Sigma^{-+}_{R,xy} S^{+-}_{R,yx}- \Sigma^{+-}_{R,xy} S^{-+}_{R,yx}  ] =S_{f}^\textsc{cp}+ S_{f}^{\CPV},
\label{Jall}
\end{align}
with $\Theta_{xy}= \Theta(x^0-y^0)$.
In the expression on the right-hand-side we split the current in a CP-conserving source $S_{i}^\textsc{cp} \equiv S_{R,i}^\textsc{cp}$ and a CP-violating source $S_{i}^{\CPV} \equiv S_{R,i}^{\CPV}$, and to avoid notational clutter we have dropped the flavor index.
The relaxation rate extracted from the CP-conserving source term for the right-handed particles is\footnote{The rescaled relaxation rate that usually appears in the transport equations found in the literature is $ \Gamma_M^\pm = \frac{6}{T^2} \Gamma^\pm$. If the fermions/scalars are (s)quarks there is an additional $N_c$ color factor.}
\be
S^\textsc{cp}_i = -\Gamma_i^+(\mu_R+\mu_L) - \Gamma_i^- (\mu_R -
\mu_L)\equiv -\Gamma_i^+ \mu^+ - \Gamma_i^- \mu^- \, ,
\label{Gamma}
\ee
where we suppressed the scalar/fermion index. \Cref{Jall,Gamma} are the master equations to calculate the source terms and rates.

We can expand the scalar self-energy in the coupling $f_s$:
 \be
 \Pi^\lambda = \Pi^\lambda_{R,\textsc{lo} }+ \Pi^\lambda_{R,\textsc{lno} } +..., \qquad
 \Sigma^\lambda = \Sigma^\lambda_{R,\textsc{lo} }+ \Sigma^\lambda_{R,\textsc{nlo} }+....
  \label{expansion}
 \ee
 The LO and NLO Feynman diagrams for the self-energy of the right-handed scalar are shown in \cref{fig:diagrams}.  Note that it is an expansion in the coupling $f_s$, that is, in the number of vev insertions, not a loop expansion.  The LO term is $ \Pi^\lambda_{R,\textsc{lo} }={\cal O}(f_s^2)$, the NLO term $ \Pi^\lambda_{R,\textsc{nlo} }={\cal O}(f_s^4)$, and the ellipses denote ${\cal O}(f_s^6)$. An analogous expansion holds for the fermionic self-energies. Consequently, we can also expand the source terms as $S^I = S^I_\textsc{lo} + S^I_\textsc{nlo}+...$ with $I ={\scriptsize \textsc{CP}},\,\CPV$.

\begin{figure}[t]
\begin{center}
\begin{tikzpicture}[line width=1.5 pt, scale=1]
		\draw(0,0) -- (0.5,1);
                \draw(0.5,1) -- (2.5,1);
                \draw(2.5,1) -- (3,0);
                \draw[dashed](0.5,1) -- (0.5,2);
                \draw[dashed](2.5,1) -- (2.5,2);
		\node at (0.5,2) {$\otimes$};
                \node at (2.5,2) {$\otimes$};	
	\node at (0.75,1.2) {$+$};	
	\node at (2.2,1.2) {$-$};
        \node at (1.5,1.3) {$\phi_L$};
	\node at (.5,0.15) {$\phi_R$};
        \node at (2.5,0.15) {$\phi_R$};
        \node at (-1.5,1) {$\Pi_R^>=$};
         \node at (4.2,1) {$+$};
\begin{scope}[shift={(5,0)}]
                \draw(0,0) -- (0.5,1);
                \draw(0.5,1) -- (6.5,1);
                \draw(6.5,1) -- (7,0);
                \draw[dashed](0.5,1) -- (0.5,2);
                \draw[dashed](2.5,1) -- (2.5,2);
                \draw[dashed](4.5,1) -- (4.5,2);
                \draw[dashed](6.5,1) -- (6.5,2);
	\node at (0.75,1.2) {$+$};	
	\node at (2.2,1.2) {$\pm$};	
        \node at (1.5,1.35) {$\phi_L$};
	\node at (2.75,1.2) {$\pm$};	
	\node at (4.2,1.2) {$\pm$};	
\node at (3.5,1.35) {$\phi_R$};
	\node at (4.75,1.2) {$\pm$};	
	\node at (6.2,1.2) {$-$};	
\node at (5.5,1.35) {$\phi_L$};
		\node at (0.5,2) {$\otimes$};
                \node at (2.5,2) {$\otimes$};	
                \node at (4.5,2) {$\otimes$};
                \node at (6.5,2) {$\otimes$};
                \node at (.5,0.15) {$\phi_R$};
                \node at (6.5,0.15) {$\phi_R$};
                    \node at (8,1) {$+\;  ...$};
\end{scope}
	 \end{tikzpicture}	
\end{center}
\caption{Feynman diagrams for the LO and NLO contributions to the self-energy $\Pi^>_R$.  The legs with $\otimes$ symbols denote vev insertions $\vp_b$, and the plus/minus signs at the vertices indicate the coupling $f_s^\pm$.  The diagrams for the fermionic self-energy $\Sigma^>_R$ have the same structure with the replacement $\phi_{L,R} \to \psi_{L,R}$. }
\label{fig:diagrams}
\end{figure}

 \subsection{Contour integral}
 \label{s:contour}

To calculate the source terms we will have to perform multiple contour integrals.  Here we introduce the master integrals, which also serve to set the notation.  The integrals are of the form
\begin{align}
  {\cal I}^\pm 
  = \int_0^{\infty} \dd u^0 \xi^I(u^0) \int_{k^0} \e^{\pm i k^0 u^0} \rho(k^0)  g(k^0) .
\end{align}
For the scalar model $g(k_0) =g^\lambda(k_0)$ the thermal distribution function \cref{gexpB}, for the fermionic model $g(k_0)$ denotes the thermal distribution function \cref{gexpF} times a $k^0$-dependent trace factor.  The function $\xi^I$  in the calculation of the CP-conserving and CP-violating source  respectively is
\be
           \xi^\textsc{cp}(x) =1, \quad 
           \xi^{\CPV}(x)  = x \, .
 \label{xi}
\ee
For ${\cal I}^+$ the contour is closed in the upper half plane where the spectral function  $\rho(k^0)$ has two poles $U(k)$. The contour for ${\cal I}^-$ is closed  in the lower half plane with poles at $D(k)$, with
\be
U(k) =\{U_1(k),U_2(k)\} = \{\E_k,-\E_k^*\},\qquad D(k) =\{D_1(k),D_2(k)\}  = \{-\E_k,\E_k^*\}.
\label{UD}
\ee
Then
\begin{align}
 {\cal I}^+ &=\int_0^{\infty} \dd u^0 \xi^I (u^0) \int_{k^0} \e^{i k^0 u^0} \rho(k^0) g(k^0)
    \hspace{0.25cm}  = - \int_0^{\infty} \dd u^0 \xi^I (u^0) \, \sum_{U} \( \frac{ (-1)^{F}\e^{i k^0 u^0} g(k^0)}{2\omega_k}\) \, ,
      \nn \\
 {\cal I}^- &=\int_0^{\infty} \dd u^0 \xi^I (u^0) \int_{k^0} \e^{-i k^0 u^0} \rho(k^0)  g(k^0)
      = \hspace{0.25cm} \int_0^{\infty} \dd u^0 \xi^I (u^0)  \, \sum_{D} \( \frac{ (-1)^{F}\e^{-i k^0 u^0} g(k^0)}{2\omega_k}\) \, ,
      \nn \\ 
\label{contour}
 \end{align}
 where we introduced the notation $\Sigma_{U} =\Sigma_{k^0 =U(k)}= \Sigma_{k^0 = \{U_1,U_2\}}$ and similarly for $\Sigma_{D}$.   Further $F = 1$ for the $k^0= \pm \E$ poles, and $F = 0$ for the $k^0= \pm\E^*$ poles, as the former pick up an extra minus sign from the residue. If there are several contour integrals over $k^0_i$ the notation is generalized to $\Sigma_{DU} = \Sigma_{k_1^0 = D}\Sigma_{k_2^0 = U}$, etc. The sign is then denoted by $(-1)^{\sum F_i}$. 
 
The final $u^0$-integral converges as $\Im(k^0)$ cuts off the integral at $u^0 \to \infty$ and
 \be
 \int_0^\infty \dd u^0 \e^{\pm ik^0 u^0}  = \pm\frac{i}{k_0},\qquad
 \int_0^\infty \dd u^0 \, u^0 \e^{\pm ik^0 u^0}  = -\frac{1}{k_0^2}.
 \label{y_integral}
 \ee

 The LO and NLO source terms require two and four contour integrals respectively. It will be useful to divide the summation over the poles into ``opposite'' halves, as these contributions either add or subtract.  For example, consider the double sum  $\Sigma_{DU}$.
 If one term in the summation is $(k_1^0,k_2^0 ) = (D_a,U_b)$ then we define the opposite term as $ (D_{\bar a},U_{\bar b})$ with $\bar a=1$ if $a =2$ and $\bar a=2$ if $a=1$. In this way we split the summation in halves, which we denote by
 \be
 \sum_{DU} = \sum_{\overline{DU}} + \sum_{\underline{DU}} \, ,
 \ee
 where the 2nd sum on the right-hand-side contains all the opposites of the terms in the first sum (which of the pair of opposites is in $ {\overline{DU}} $ is arbitrary).  We can then use the relations $U_a = - U_{\bar a}^* $ and similar for $D$ to simplify relations.

\section{Leading order source terms}
\label{s:LO}

In this section we review the LO calculation of the CP-conserving and -violating source terms appearing in the transport equation for the right-handed fermion/scalar. 

\paragraph{Scalars}

At leading order the self-energy of the right-handed scalar, given by the Feynman diagram in \cref{fig:diagrams}, is
\be
\Pi^{\lambda}_{R,\textsc{lo}}(x,y)= -(-if_x^-) (-if_y^{+*})  G_{L,xy}^{\lambda} 
= - f_xf^*_y   G_{L,xy}^{\lambda}\, ,
\label{PiR}
\ee
 where we used the notation $G_{L,xy}^{\lambda} = G_{L}^{\lambda}  (x,y)$, $f_x=f(x)$ etc,  and $\lambda \in \{>,\, <\}$. We suppressed the label for scalars on $f_s$. To get the final expression we used \cref{sign_f}.  As $f_xf^*_y = (f_y f^*_x)^*$, and the Green's functions obey \cref{Gdagger}, it follows that \cref{Pidagger} is satisfied as well.  From \cref{Jall}, the transport equation at LO is
\be
\partial_\mu j_{R}^\mu= S_\textsc{lo}^\textsc{cp}+ S_\textsc{lo}^{\CPV} = -2\int \dd^4 y \, \Theta_{xy}\, \Re\[ f_xf^*_y\(G_{L,xy}^{>}  G_{R,yx}^{<}
                -  G_{L,xy}^{<}   G_{R,yx}^{>} \)\] \, ,
\ee
with
\begin{align}
S_\textsc{lo}^{\textsc{cp}} 
&= 
 -2\int \dd^4 y \, \Theta_{xy}\,
\Re[f_xf^*_y ]\Re\[G_{L,xy}^{>}  G_{R,yx}^{<}
                -  G_{L,xy}^{<}   G_{R,yx}^{>}\],\nn \\
S_\textsc{lo}^{\CPV} &=\;\;\;2\int
 \dd^4 y \, \Theta_{xy}\,
\Im[f_xf^*_y]\Im \[G_{L,xy}^{>}  G_{R,yx}^{<}
             -  G_{L,xy}^{<}   G_{R,yx}^{>}\].
             \label{J_LO_B}
\end{align}

\paragraph{Fermions}

The self-energy for the right-handed fermion at leading order is

\be
\Sigma^\lambda_{R,\textsc{lo}}(xy) = - \frac12 f_x f_y^* P_R S_{L,xy}^\lambda P_L
\label{SigmaR}
\ee
with $P_{L,R}$ the left- and right-handed projection operators.
Substituting in the transport \cref{Jall}, and inserting the propagators \cref{Sk}, the trace is of the form
\be
{\rm Tr} \[ P_L (\slashed{k_1} + m_1) P_R (\slashed{k_2} + m_2)\]
={\rm Tr} \[ P_L \slashed{k_1} P_R \slashed{k_2} \] 
=\frac12 {\rm Tr} \[ \slashed{k_1} \slashed{k_2} \].
\label{trace}
\ee
Hence, we can neglect the mass-term in the trace and work with the propagators defined in \cref{SKtrace}.  Following the same steps as for the scalar case, the result for the CP and CPV source is
\begin{align}
S_\textsc{lo}^{\textsc{cp}} 
&= \;\;\;
 \int \dd^4y \,\Theta_{xy}\,
\Re[f_xf^*_y]  \Re \Tr\[S_{L,xy}^{>}  S_{R,yx}^{<}
                -  S_{L,xy}^{<}   S_{R,yx}^{>}\]_{\Tr(m)=0},\nn \\
S_\textsc{lo}^{\CPV} &= -
 \int \dd^4y \,\Theta_{xy}\,
\Im[f_xf^*_y]\Im \Tr \[S_{L,xy}^{>}  S_{R,yx}^{<}
             -  S_{L,xy}^{<}   S_{R,yx}^{>}\]_{\Tr(m)=0}.
               \label{J_LO_F}
\end{align}
%

\subsection{Derivative expansion}

Let's start with the CP-conserving source term. At leading order in the derivative expansion, valid when the bubble wall background changes slowly compared to timescales set by the temperature, we can approximate $\Re[f_xf^*_y] \approx |f(x)|^2$ and take it out of the integral for the CP-even source in \cref{J_LO_B,J_LO_F}.  Now insert the explicit form of the propagators \cref{Gk,SKtrace}.  The integration over spatial coordinates can readily be done, and gives a delta function that sets all spatial momenta equal.  We further introduce a new time coordinate $u = x^0-y^0$, such that the theta function becomes $\Theta(u)$ and we only have to integrate over positive time values.  Then
\be
S_\textsc{lo}^\textsc{cp} 
= -2s |f|^2 \Re \int_0^{\infty} \dd u \int_{k_1} \int_{k_2^0}
\e^{-i (k_1^0-k_2^0) u } c^{+-}(k_i^0)  \rho_L(k_1) \rho_R(k_2) \tr^\textsc{lo} (k_i) \Big|_{\vec k_i = \vec k} \, ,
\label{SRCP}
\ee
with
\be
c^{+-}(k_i^0) \equiv c^{+-}(k_1^0,k_2^0) 
= g^>(k_1^0,\mu_L) g^<(k_2^0,  \mu_R)- g^<(k_1^0, \mu_L) g^>(k_2^0,  \mu_R).
\label{c_pm}
\ee
The momentum $k^0_1$ corresponds to the left-handed particle with chemical potential $\mu_L$, the momentum $k^0_2$ to the right-handed particle with chemical potential $\mu_R$.  \Cref{SRCP} is valid for scalars as well as fermions.  For scalars $s=1$ and the trace factor is trivial $\tr^\textsc{lo} (k_i) =1$, $f= f_s$, and the distribution functions $g^\lambda$ depend on the Bose-Einstein distributions \cref{gexpB}.  For fermions $s=-1 $,
\be
\tr^\textsc{lo} (k_i) = \frac14\Tr(\slashed{k_1}\slashed{k_2}) =(k_1^0 k_2^0 -\vec k^2).
\label{tr_lo}
\ee
 $f= f_f$,  and the distribution functions $g^\lambda$ depend on  the Fermi-Dirac distributions \cref{gexpF}.

To calculate the CP-violating source term in \cref{J_LO_B,J_LO_F} we expand the coupling 
\be
\lim_{y\to x} \(f_xf^*_y - f_yf^*_x \)
= 2i \Im\( f(x)\partial_\mu f(x)^*\)  (y-x)^\mu
\equiv  2i (y-x)^0 \delta \, ,
\ee
where in the last step we only included the time-derivative term as the spatial part cancels in the source integral because of spatial isometry, and we defined $\delta=\Im\big( f(x)\dot f(x)^*\big)$.
The source term becomes
\begin{align}
S_\textsc{lo}^{\CPV} 
&= -2s\delta \,\Im  \int_0^{\infty} \dd u \,u \int_{k_1}\int_{k_2^0}\,
              \e^{-i (k_1^0-k_2^0) u } c^{+-}(k_i^0)  \rho_L(k_1) \rho_R(k_2) \tr^\textsc{lo}  (k_i) \Big|_{\vec k = \vec k_i},
            \label{SRCPV}
\end{align}
which again applies to both scalars and fermions. 

\subsection{Contour/$k^0_i$ Integrals}

The integrals in the CP-conserving and violating source terms \cref{SRCP,SRCPV}
can be written as  $S_\textsc{lo}^\textsc{cp} =-2s |f|^2 \Re \int_{\vec k} {\cal J}_\textsc{lo}^\textsc{cp}$ and $S_\textsc{lo}^{\CPV} =-2s\delta \,\Im \int_{\vec k} {\cal J}_\textsc{lo}^{\CPV}$, with
\be
{\cal J}_\textsc{lo}^I
=  \int_0^{\infty} \dd u \,\xi^I(u)\int_{k_1^0} \int_{k_2^0} 
\e^{-i (k_1^0-k_2^0) u } c^{+-} (k_i^0)  \rho_L(k_1) \rho_R(k_2) \tr^\textsc{lo} (k_i) \Big|_{\vec k = \vec k_i} \, ,
\ee
and $\xi^\textsc{cp}=1$ and $\xi^{\CPV}(u)=u$ as in \cref{xi}.
To do the contour integrals we close the contour for $k_1^0$ below and for $k_2^0 $ above the real axis.  There are two poles per integral, giving rise to four residue terms in total. Using \cref{contour,y_integral} the result is
\begin{align}
{\cal J}^I_\textsc{lo} &= -\frac{1}{4\omega_L  \omega_R} \int \dd u \, \xi^I(u) \sum_{DU}\[ (-1)^{\sum F_i}\e^{-i (k_1^0 -k_2^0) u} c^{+-} \tr^\textsc{lo}\]\nn\\
                  &= \frac{1 }{4\omega_L  \omega_R}
                    \sum_{DU}i\[ \frac{ (-1)^{\sum F_i} c^{+-} \tr^\textsc{lo}(k_i)}{k_1^0-k_2^0}\Xi^I(k^0_i) \] \, ,
  \label{J_LO}
\end{align}
with 
\be
\Xi^\textsc{cp}(k^0_i) =1,\qquad
\Xi^{\CPV}(k^0_i) = 
                        \frac{-i}{k_1^0-k_2^0} \, .
\ee                    

Now expand $c^{+-}= c^{+-}_{0} + c^{+-}_{1}+ {\cal O  }(\mu_i^2) $  in small chemical potentials using \cref{gexpB,gexpF}, with the 0th order term $c^{+-}_{0}$  independent of $\mu_i$ and the the 1st order term $c^{+-}_{1}$ linear in $\mu_i$.
Using the Bose-Einstein and Fermi-Dirac distributions for scalars and fermions respectively, we get
\begin{align}
  c^{+-}_{0} (k_1^0,k_2^0)&
                            =- s(n(k^0_1) -n(k^0_2)), \nn \\
  c^{+-}_{1}(k_1^0,k_2^0)   & 
                   = \frac{s}{2T}\[ \mu^+\(h(k^0_1)  -h(k^0_2)\) -\mu^-\(h(k^0_1)  +h(k^0_2)\)\],
  \label{c0}
\end{align}
 with $s=1$ for scalars and $s=-1$ for fermions, and 
 $\mu^\pm=\mu_R\pm\mu_L$.   The summation in \cref{J_LO} can be divided into the poles $(a,b)\in \overline{DU}$ and the opposite pairs $(\bar a,\bar b) \in \underline{DU}$, as discussed in \cref{s:contour}, where we introduced the notation that $(a,b)$ denotes $k^0_1 =D_{a},\, k^0_2 = U_{b}$.
 Then
\begin{align}
  c^{+-}_{0}\big|_{(\bar a,\bar b)}= -(c^{+-}_{0})^*\big|_{( a, b)},\qquad
  c^{+-}_{1}\big|_{(\bar a,\bar b)}= (c^{+-}_{1})^*\big|_{(a, b)} \, ,
\end{align}
where we used $g^> (-k^0,0) = - g^<(k^0,0)$ and $h(-k^0) =h(k^0)$.  This can be used to simplify ${\cal J}^I_\textsc{lo}$.

\subsection{Relaxation rate}

For the CP-conserving source it follows that the leading order term in the $\mu$-expansion cancels.  The summation in \cref{J_LO} is over $A_0 = (-1)^{\sum F_i} c^{+-}_{0} \tr^\textsc{lo}/(k_1^0-k_2^0)$ with $A_0|_{(\bar a,\bar b)}= A_0^*|_{(a,b)}$, and thus
\be
  \Re \big[i\sum_{DU}  A_0 \big]
= \Re\big[i \big( \sum_{\overline{DU}}
                A_0 +\sum_{\underline{DU}}
                 A_0 \big)\big]
  =\Re\big[i \sum_{\overline{DU}}
                (A_0+A_0^*) \big]=0 \, .
                \label{cancel_LO}
  \ee
At first order in the chemical potential the summation is over $A_1 = (-1)^{\sum F_i} c^{+-}_{1}  \tr^\textsc{lo}/(k_1^0-k_2^0)$ with $A_1|_{(\bar a,\bar b)}= -A_1^*|_{(a,b)}$, and we get instead
\be
  \Re \big[ i\sum_{DU} A_1 \big]
=\Re \big[i\sum_{\overline{DU}} \(
                 A_1 - A_1^*\)\big]
                =-2  \Im \big[\sum_{\overline{DU}}   A_1 \big].
                \label{add_LO}
\ee
Putting it all together, we find for the CP conserving source (setting $k_1^0 =k_L^0$ and $k_2^0 =k_R^0$ for the left- and right-handed particle respectively)
\begin{align}
  S_\textsc{lo}^\textsc{cp} &=-2s |f|^2 \int_{\vec k} \Re[{\cal J}_\textsc{lo}]
                  = s|f|^2 \Im \int_{\vec k}\frac{1}{\omega_L  \omega_R}
                    \sum_{\overline{DU}}\bigg[ \frac{ (-1)^{\sum F_i} c^{+-}_{1} (k^0_i) \tr^\textsc{lo}(k_i)}{k_L^0-k_R^0}\bigg] \, ,
 \end{align}
from which we extract the relaxation rates (see \cref{Gamma})
 \begin{align}
\Gamma^\pm_\textsc{lo}  &= \mp \frac{|f|^2}{2T}\Im \int_{\vec k } \frac1{\omega_L
                    \omega_R} \sum_{\overline{DU}}\bigg[ \frac{ (-1)^{\sum F_i} \(h(k^0_L)  \mp h(k^0_R)\)\tr^\textsc{lo}}{k_L^0-k_R^0}\bigg].
\end{align}
As the final step,  we sum over the poles in $\overline{DU}$ which we choose $(k_1^0,k_2^0) =(\E_L^*,\E_R),\ (-\E_L,\E_R)$, where for the first combination $(-1)^{\sum F_i} =-1$ and for the second $(-1)^{\sum F_i} =1$. The relaxation rate becomes 
\begin{align}
  \Gamma_\textsc{lo}^\pm 
    &= \pm \frac{|f|^2}{2T} \Im \, \int_{\vec k}\frac{1}{\omega_L  \omega_R}
 \[\frac{\(h(\E_L^*) \mp h(\E_R) \)}{ \E_L^*-\E_R}\tr^\textsc{lo}_1
   + \frac{\( h(\E_L) \mp  h(\E_R) \)}{\E_L +\E_R}\tr^\textsc{lo}_2    
    \] \, ,
    \label{gamma0}
\end{align}
with for scalars $\tr^\textsc{lo}_i =1$, and for fermions
\begin{align}
  \tr^\textsc{lo}_1 =  \tr^\textsc{lo} \big|_{\{k_1^0= \E_L^*,k_2^0 = \E_R\}} =(\E_L^* \E_R -\vec k^2) ,
  \qquad
  \tr^\textsc{lo}_2 =  \tr^\textsc{lo} \big|_{\{k_1^0= -\E_L,k_2^0 = \E_R \}}= -(\E_L \E_R +\vec k^2)\, ,
\label{tr0}
\end{align}
with $\tr^\textsc{lo}$ given in \cref{tr_lo}.

\subsection{CP-violating source}

Because of the non-trivial $\Xi^{\CPV}$-factor in \cref{J_LO} for  the CP-violating source, the leading order term in the $\mu$-expansion already contributes.  Indeed $(iA_0 \Xi^{\CPV})|_{(\bar a,\bar b)} =( iA_0 \Xi^{\CPV})^*|_{( a,b)}$ and
\begin{align}
  \Im \big[i \sum_{DU}  A_0 \Xi^{\CPV} \big]
&= \Im \big[  \sum_{\overline{DU}} \big(
               (iA_0 \Xi^{\CPV})-
                 (i A_0 \Xi^{\CPV})^* \big) \big]
                =2  \Im \big[  \sum_{\overline{DU}}   ( i A_0\Xi^{\CPV})\big] \, .
\end{align}
Putting it all together, then
\begin{align}
  S_\textsc{lo}^{\CPV} &=-2s\delta \,\Im \int_{\vec k} [{\cal J}^{\CPV}_\textsc{lo}]
                   = -s\delta \,\Im \int_{\vec k} \frac1{\omega_L  \omega_R}
                    \sum_{\overline{DU}}\bigg[ \frac{ (-1)^{\sum F_i} c^{+-}_{0} \tr^\textsc{lo}(k_i)}{(k_1^0-k_2^0)^2} \bigg]
\nn\\
           &= \delta \,\Im \int_{\vec k} \frac1{\omega_L  \omega_R}
             \sum_{\overline{DU}} \bigg[ \frac{(-1)^{\sum F_i}(n(k_L^0)-n(k_R^0))\tr^\textsc{lo}(k_i)}{(k^0_L-k^0_R)^2}\bigg]    \, .
\end{align}
Choosing to sum over $(k_1^0,k_2^0) = (\E_L^*,\E_R),\,(-\E_L,\E_R)$ we get the final result
\begin{align}
S_\textsc{lo}^{\CPV} 
           &=-\delta\,\Im  \int_{\vec k}\frac{1}{\omega_L  \omega_R}
         \[ \frac{n(\E_L^*) -n(\E_R)}{(\E_L^*-\E_R)^2}\tr_1^\textsc{lo}
             +\frac{n(\E_L) + n(\E_R)+s}{(\E_L+\E_R)^2}\tr_2^\textsc{lo}  
 \] \, ,
\end{align}
with $\tr_i^\textsc{lo}$ defined in \cref{tr0}. In the second term between square brackets, the term $\propto s$ in the numerator diverges, also in the zero-temperature limit; it can be removed by adding a counterterm and is absent in the renormalized theory \cite{Liu:2011jh}.

\section{Next-to-leading order source terms}
\label{s:NLO}

In this section we calculate the source terms in the transport equation \cref{Jall} at NLO. The self-energy at this order is given by the second Feynman diagram in \cref{fig:diagrams}.  It is still a tree level diagram, but now with four mass insertions.  The NLO diagram thus represents the next order contribution in the coupling expansion, and is ${\cal O}( |f|^2)$ suppressed with respect to the LO diagram.  

Although the extra mass insertions complicate the calculation of the relevant integrals, since the self-energy diagram remains tree level, all steps in the calculation are essentially the same as for the LO calculation.

\paragraph{Scalars}

Let's start with the scalar set-up.  The self-energy of the right-handed scalar at NLO is
\begin{align}
  \Pi^{-+}_{R,\textsc{nlo}}  (x,y)&=-\sum_{a_1,a_2=\pm}\int \dd^4 z_1 \dd^4 z_2 \, \(f^-_x f_{z_1}^{a_1 *} f_{z_2}^{a_2} f^{+*}_y\) \,
 G_{L,xz_1}^{-a_1} G_{R,z_1 z_2}^{a_1a_2} G_{L,z_2 y}^{a_2 +} \nn \\
                 &= \int \dd^4 z_1 \dd^4 z_2 \, \(f_x f_{z_1}^* f_{z_2} f^{*}_y\)              
                   \big[ G_{L,xz_1}^{-+} G_{R,z_1 z_2}^{++} G_{L,z_2 y}^{++} 
+G_{L,xz_1}^{--} G_{R,z_1 z_2}^{--} G_{L,z_2 y}^{-+} 
\nn \\ & \hspace{4.5cm}
           - G_{L,xz_1}^{-+} G_{R,z_1 z_2}^{+-} G_{L,z_2 y}^{-+} 
         - G_{L,xz_1}^{--} G_{R,z_1 z_2}^{-+} G_{L,z_2 y}^{++}  \big] \, ,
         \label{PiNLO}
\end{align}
 where we summed over all possible $\pm$-signs for the internal vertices. Using \cref{Gdagger}  it follows that $( \Pi^{-+}_{R,\textsc{nlo}}(x,y) )^\dagger   =\Pi^{-+}_{R,\textsc{nlo}}(y,x)$.
We can then extract the CP-conserving and -violating source terms from \cref{Jall} 
\begin{align}
  S_\textsc{nlo}^\textsc{cp} & =\;\;\;2 \int d^4 y \int \dd^4 z_1 \dd^4 z_2 \, \Theta_{xy}\,
                \Re[f_x f_{z_1}^* f_{z_2} f^{*}_y ] \,\Re(\hat I_1+\hat I_2),
 \nn \\                
 S_\textsc{nlo}^{\CPV} & = -2\int d^4 y \int \dd^4 z_1 \dd^4 z_2 \, \Theta_{xy}\,
              \Im [f_x f_{z_1}^* f_{z_2} f^{*}_y] \,\Im(\hat I_1+\hat I_2),
              \label{J_NLO_B}
\end{align}
with 
\begin{align}
 \hat I_1
    &=\big[G_{L,xz_1}^{-+}  G_{R,z_1 z_2}^{++} G_{L,z_2 y}^{++} 
+ G_{L,xz_1}^{--} G_{R,z_1 z_2}^{--} G_{L,z_2 y}^{-+} 
\nn\\ &\hspace{1cm}
- G_{L,xz_1}^{-+} G_{R,z_1 z_2}^{+-} G_{L,z_2 y}^{-+} 
        - G_{L,xz_1}^{--}G_{R,z_1 z_2}^{-+} G_{L,z_2 y}^{++} \big] G^{+-}_{R,yx} \, ,
        \nn\\
\hat I_2
    &= -\big[G_{L,xz_1}^{++}  G_{R,z_1 z_2}^{++} G_{L,z_2 y}^{+-} 
+ G_{L,xz_1}^{+-} G_{R,z_1 z_2}^{--} G_{L,z_2 y}^{--} 
\nn\\ &\hspace{1cm}
- G_{L,xz_1}^{++} G_{R,z_1 z_2}^{+-} G_{L,z_2 y}^{--} 
        - G_{L,xz_1}^{+-}G_{R,z_1 z_2}^{-+} G_{L,z_2 y}^{+-} \big] G^{-+}_{R,yx} \, .
       \label{hatI}
\end{align}

\paragraph{Fermions}

The self-energy for the right-handed fermion at NLO is
\begin{align}
\Sigma^{-+}_{R,\textsc{nlo}}  (x,y)&=-\frac14\sum_{a_1,a_2=\pm}\int \dd^4 z_1 \dd^4 z_2 \, \(f^-_x f_{z_1}^{a_1 *} f_{z_2}^{a_2} f^{+*}_y\) \,
                      P_R S_{L,xz_1}^{-a_1} P_L S_{R,z_1 z_2}^{a_1a_2} P_R S_{L,z_2 y}^{a_2 +} P_L,
                     \end{align}
which has the property that $ \(\Sigma^{-+}_{R,\textsc{nlo}}(x,y) \)^\dagger =\gamma^0 \Sigma^{-+}_{R,\textsc{nlo} }(y,x)\gamma^0$.  When used in the transport equation \cref{Jfermion}, the trace of the $\Sigma S$ and $S \Sigma$-terms is respectively
\begin{align}
{\rm Tr} \[ P_R (\slashed{k_1} + m) P_L (\slashed{k_2} + m) P_R (\slashed{k_3} + m) P_L (\slashed{k_4} + m)
\] &= \Tr\[P_R\slashed{k_1}\slashed{k_2}\slashed{k_3}\slashed{k_4}\] \rightarrow \frac12 \Tr\[\slashed{k_1}\slashed{k_2}\slashed{k_3}\slashed{k_4}\]\, ,\nn \\
{\rm Tr} \[ (\slashed{k_1} + m) P_R (\slashed{k_2} + m) P_L (\slashed{k_3} + m) P_R (\slashed{k_4} + m)P_L
\] &= \Tr\[P_L\slashed{k_1}\slashed{k_2}\slashed{k_3}\slashed{k_4}\] \rightarrow \frac12 \Tr\[\slashed{k_1}\slashed{k_2}\slashed{k_3}\slashed{k_4}\]\, .
\end{align}
The last expression holds since the spatial integration gives a set of delta functions that set $\vec k_i = \vec k_j$. We can then use that $\Tr\[\gamma^5\slashed{k_1}\slashed{k_2}\slashed{k_3}\slashed{k_4}\]=-2\eps_{\mu \rho \kappa \sigma} k_1^\mu k_2^\rho k_3^\kappa k_4^\sigma =0$.

The source terms for fermions are then analogous to those for the scalars
\begin{align}
S_\textsc{nlo}^{\textsc{cp}} 
&= - \frac14
\int d^4 y \int \dd^4 z_1 \dd^4 z_2 \, \Theta_{xy}\,  \Re[f_x f_{z_1}^* f_{z_2} f^{*}_y]\, \Re \, \Tr [\hat I_1 +\hat I_2]_{\Tr(m)=0}\, ,\nn \\
S_\textsc{nlo}^{\CPV} &= \;\;\; \frac14\int d^4 y \int \dd^4 z_1 \dd^4 z_2 \, \Theta_{xy}\,
             \Im[f_x f_{z_1}^* f_{z_2} f^{*}_y ]\,\Im \, \Tr [\hat I_1 +\hat I_2]_{\Tr(m)=0} \, ,
              \label{J_NLO_F}
\end{align}
with $\hat I_1,\hat I_2$ as in \cref{hatI} with the replacement $G_{xy} \to S_{xy}$.

\subsection{Derivative expansion}

Let's start again with the CP-conserving source.  To lowest order in the derivative expansion $f_x f_{z_1}^* f_{z_2} f^{*}_y\approx |f(x)|^4$, and it can be taken out of the integrals for the CP even source in \cref{J_NLO_B,J_NLO_F}.  Now we express all time and anti-time ordered propagators in terms of the Wightman functions using \cref{Gplus}, and insert the explicit expressions for the latter \cref{Gk,SKtrace}.  The integration over spatial momenta and coordinates sets all spatial momenta equal.  The CP-even source becomes
\begin{align}
  S_{\textsc{nlo}}^{\textsc{cp}}
   & =2s|f|^4 \Re \int \dd z^0_1 \dd z^0_2 \dd z_3^0 \int_{\vec k} \; \int_{k^0_1,k^0_2,k^0_3,k^0_4} \,  
                   \rho_{L}(k_1)     \rho_{R}(k_2)     \rho_{L}(k_3)     \rho_{R}(k_4)   \label{ac1}
\\ & \times
        \e^{-ik_1^0(x^0-z_1^0) -ik_2^0(z_1^0-z_2^0)-ik_3^0(z_2^0-z_3^0)-ik_4^0(z_3^0-x^0)} \,
       \sum_{i_1,i_2,i_3,i_4=\pm} \!\!\!\!  \theta_{i_1 i_2 i_3 i_4} (z_i) \, c_{i_1 i_2i_3 i_4}(k_i^0) \, \tr^\textsc{nlo} \Big|_{\vec k_i=\vec k} \,  \nn \, .
\end{align}
The particles with momenta $k_1,k_3$ are left-handed, those with momenta $k_2,k_4$ right-handed.
Here $c_{i_1 i_2i_3 i_4}$ denotes the combination of distribution functions
  \be
  c_{i_1 i_2i_3 i_4}= g^{i_1}_{L}(k_1^0)   g^{i_2}_{R}(k_2^0)  g^{i_3}_{L}(k_3^0) g^{i_4}_{R}(k_4^0)  - g^{\bar i_1}_{L}(k_1^0)   g^{\bar i_2}_{R}(k_2^0)  g^{\bar i_3}_{L}(k_3^0) g^{\bar i_4}_{R}(k_4^0),
  \label{def_c}
  \ee
  with $i_j = \pm$ and $\bar i_j$ is  $+$ if $i_j$ is $-$, and vice versa. $g_{L,R}^{i_j}(k^0_i) = g_{L,R}^>(k^0_i)$ when $i_j = +$ and $g_{L,R}^{i_j}(k^0_i) = g_{L,R}^<(k^0_i)$ when $i_j = -$ . Further $\theta_{i_1 i_2i_3 i_4}$ are the time-dependent coefficients, which are a combination of $\Theta$-functions
\begin{align}
 \theta_{++++} &= 0, &
  \theta_{+---} &= \Theta_{xz_3}\Theta_{z_3 z_2} \Theta_{z_2 z_1}, \label{a_values}                \\
\theta_{-+--}   &=    - \Theta_{xz_3}\Theta_{x z_1} \Theta_{z_3 z_2}, & 
 \theta_{++--}    &=   \Theta_{xz_3}  \(\Theta_{z_3 z_2} \Theta_{z_1 z_2} -\Theta_{z_3 z_2} \Theta_{z_1 x}\),     \nn \\  
\theta_{--+-}      &= \Theta_{xz_3}\Theta_{x z_1} \Theta_{z_1 z_2}, &
 \theta_{+-+-}    &= \Theta_{xz_3}\( \Theta_{z_1 x} \Theta_{z_1 z_2}+\Theta_{z_2 z_3} \Theta_{z_2 z_1}-1\) ,    \nn \\ 
\theta_{-++-} &=\Theta_{xz_3}  \( \Theta_{x z_1} \Theta_{z_2 z_1}-\Theta_{x z_1} \Theta_{z_2 z_3}\) ,&  
                                                                                                   \theta_{+++-}  &=\Theta_{xz_3} \(\Theta_{z_1 z_2} \Theta_{z_2 z_3}+\Theta_{z_1 x} \Theta_{z_2 z_1}-\Theta_{z_1 x} \Theta_{z_2 z_3}\). \nn                                                                                              
\end{align}
For scalars $s=1$ and $\tr^\textsc{nlo} =1$, while for fermions $s=-1$ and the trace factor now becomes
\be
  \tr^\textsc{nlo} \equiv  \frac18\Tr  \[\slashed{k_1}\slashed{k_2} \slashed{k_3}\slashed{k_4}\] = \frac12\[ (k_1 \cdot k_2)( k_3 \cdot k_4) -( k_1 \cdot k_3)( k_2 \cdot k_4)+(k_1 \cdot k_4) (k_2 \cdot k_3)\],
\label{tr1}
\ee
with $k_i \cdot k_j  = k^0_i k^0_j - \vec k^2$.

For the CP-violating source we expand 
\be
\lim_{y,z_1,z_2 \to x}  \Im [f_x f_{z_1}^* f_{z_2} f^{*}_y] \simeq - \delta|f|^2 (x-y-z_1+z_2)^0\, .
\ee
Integrating over spatial momenta, the CPV source terms in \cref{J_NLO_B,J_NLO_F} are
\begin{align}
  S_\textsc{nlo}^{\CPV}
   & =2s|f|^2 \delta\, \Im \, \int \dd z^0_1 \dd z^0_2 \dd z_3^0 \, (x-y-z_1+z_2)^0 \;\int_{\vec k} \; \int_{k^0_1,k^0_2,k^0_3,k^0_4} \,    \,  
                   \rho_{L}(k_1)     \rho_{R}(k_2)     \rho_{L}(k_3)     \rho_{R}(k_4)    
\nn\\ & \times
        \e^{-ik_1^0(x^0-z_1^0) -ik_2^0(z_1^0-z_2^0)-ik_3^0(z_2^0-z_3^0)-ik_4^0(z_3^0-x^0)} \,
      \!\!\!\! \!\!  \sum_{i_1,i_2,i_3,i_4=\pm} \!\! \!\!\!\! \theta_{i_1 i_2 i_3 i_4} (z_i) \, c_{i_1 i_2i_3 i_4}(k_i^0) \, \tr^\textsc{nlo} \Big|_{\vec k=\vec k_i} \, ,
        \label{ac2}
\end{align}
and $\tr^\textsc{nlo},\,\theta_{i_1 i_2 i_3 i_4},\,c_{i_1 i_2i_3 i_4}$ the same as for the CP-even source given in \cref{tr1,a_values,def_c}.

\subsection{Contour/$k^0_i$ Integrals}

The integrals in the CP-conserving and -violating source terms \cref{ac1,ac2}
can be written as $S_\textsc{nlo}^\textsc{cp}=2s|f|^4 \Re \int_{\vec k}  {\cal J}^\textsc{cp}_\textsc{nlo}$, and $S_\textsc{nlo}^{\CPV}=2s|f|^2 \delta \Im \int_{\vec k}  {\cal J}^{\CPV}_\textsc{nlo}$, with
\begin{align}
  {\cal J}_\textsc{nlo}^I
   & =\int \dd z^0_1 \dd z^0_2 \dd z_3^0  \, \xi^I( x ^0-z ^0_3-z ^0_1+z ^0_2)\int_{k^0_1,k_2^0,k_3^0,k_4^0}  \,  
                   \rho_{L}(k_1)     \rho_{R}(k_2)     \rho_{L}(k_3)     \rho_{R}(k_4)   
\nn\\ & \;\;\;\; \times
        \e^{-ik_1^0(x^0-z_1^0) -ik_2^0(z_1^0-z_2^0)-ik_3^0(z_2^0-z_3^0)-ik_4^0(z_3^0-x^0)} \,
       \!\!\!\! \sum_{i_1,i_2,i_3,i_4=\pm} \!\!\!\! \theta_{i_1 i_2 i_3 i_4} (z_i) \, c_{i_1 i_2i_3 i_4}(k_i^0) \, \tr^\textsc{nlo} \Big|_{\vec k_i=\vec k} \nn \\
   & \equiv \sum {\cal J}^I_{i_1i_2i_3i_4}.
     \label{J1}
\end{align}

We will work out two example terms $ {\cal J}^I_{i_1i_2i_3i_4}$ in \cref{J1} explicitly; all other terms can be evaluated in a similar way, and the final result is in \cref{J1res} below.
Let's start with the ${\cal J}^I_{+---} $ integral, which is relatively straightforward to integrate. First, define new time coordinates $u=x^0-z^0_3, \; v =z^0_2-z^0_1,\; w=z^0_3-z^0_2 $ such that the theta-functions become $\Theta(u) \Theta(v) \Theta(w)$, and the integration is only over positive times. The contour integrals over $k^0_i$ can now be done, closing the $k^0_2,k^0_3,k^0_4$ contour in the upper half plane with poles $k^0_i \in U(k_i^0) = \{\E_{k_i},-\E_{k_i}^*\}$, and $k^0_1$ in the lower half plane with poles $k^0_1 \in D(k_1^0) = \{-\E_{k_1},\E_{k_1}^*\}$.  The contour integral then has an overal minus sign.
  \begin{align}
&  {\cal J}^I_{+---}  \nn \\  
& \quad
                 = \int_0^\infty \dd u \dd v \dd w \,\xi^I(u+v) \int_{k^0_1,k^0_2,k^0_3,k^0_4}  \, 
                \rho_1    \rho_2     \rho_3     \rho_4 \,
   \e^{i k^0_4 u+i k^0_2 v+i k^0_3 w-i k^0_1 (u+v+w)}
   c_{+---} \,\tr^\textsc{nlo}\nn \\
  &\quad =- \,\sum_{DUUU} 
         {\cal F} c_{+---} 
  \int \dd u \dd v \dd w  \, \xi^I(u+w)  \, \e^{i u(k^0_4 -k^0_1)+i v(k^0_2 -k^0_1)+i w (k^0_3 - k^0_1)}
       \nn\\ & \quad
 = \,\sum_{DUUU} 
               \frac{i   {\cal F} c_{+---}\, \Xi^I_{DUUU} }{(k^0_2 - k^0_1) (k^0_3 - k^0_1) (k^0_4 - k^0_1)} \, ,
               \label{Jpmmm}
\end{align}
with
\be
 {\cal F} = \frac{(-1)^{\sum F_i}}{2^4 \omega_L^2 \omega_R^2} \tr^\textsc{nlo}\, .
\label{F}
\ee
For the CP even integral $\xi^\textsc{cp} =\Xi^\textsc{cp}_{DUUU} =1$, for the CP odd integral $\Xi^{\CPV}_{DUUU} $ is given in \cref{Xi} below.
In the last line of \cref{Jpmmm} we did the $u,v,w$ integration using \cref{y_integral}.  For the summation over poles we used the notation defined in \cref{s:contour}.

To evaluate all terms in \cref{J1} it is sometimes necessary to split the integration interval using theta-functions, and to do further coordinate transformations. Consider  ${\cal J}^I_{++--} $, which consists of two terms, see \cref{a_values}. To do the contour integral for the first we use $1 = \Theta(u-v+w)+\Theta(-u+v-w)$ to split the $k^0_1$ integral; the result is
\begin{align}
  &{\cal J}^I_{++--} \nn \\
& \quad =-\sum_{UDUU} 
          {\cal F} c_{++--}  
  \int_0^\infty \dd u \dd v \dd w  \, \xi^I(u-v)\e^{-i k^0_1 (u-v+w)-i k^0_2 v+i k^0_3 w+i k^0_4 u}\Theta(-u+v-w)
                          \nn\\
  & \quad \;\;\;\;+\, \sum_{DDUU} 
         {\cal F} c_{++--}  
    \int_0^\infty \dd u \dd v \dd w  \, \xi^I(u-v)\e^{-i k^0_1 (u-v+w)-i k^0_2 v+i k^0_3 w+i k^0_4 u}\Theta(u-v+w)
       \nn\\
  & \quad \;\;\;\;+\, \sum_{UDUU} 
          {\cal F} c_{++--}  
    \int_0^\infty\dd u \dd v \dd w  \, \xi^I(-v-w)\e^{i k^0_1 v-i k^0_2 (u+v+w)+i k^0_3 w+i k^0_4 u} \, .
    \label{A4cancel}
\end{align}
In the first term we can make the transformation $v'=v-u-w$ 
\begin{align}
  &\int_{-\infty}^{+\infty}\dd u \dd v \dd w\, \Theta(u)\Theta(v) \Theta(w) \Theta(v-u-w) \xi^I(u-v) f(u,v,w)
  \nn \\
    &\hspace{1cm}=
\int_{-\infty}^{+\infty} \dd u \dd v' \dd w\,\Theta(u) \Theta(w) \Theta(v') \xi^I(-v'-w) f(u,v',w) \, ,
\end{align}
and it follows that the first and third term in \cref{A4cancel} cancel.  To calculate the remaining term, use $1 = \Theta(w-v)+ \Theta(v-w)$.  In the first term do the coordinate transformation $u'=u, v'=v,w'=w-v$ and in the second use $u=u'+v',v=v'+w',w=w'$. Then
\begin{align}
  {\cal J}^I_{++--}
  &=\sum_{DDUU} 
     {\cal F} c_{++--}   \int_0^\infty \dd u' \dd v' \dd w'  \,  \Big(
    \xi^I(u'-v')\e^{i k^0_4 u'-i k^0_2 v'+i k^0_3 (w'+v')-i k^0_1 (u'+w')}
     \\ 
    &\hspace{3cm}+  \xi^I(u'-w')\e^{i k^0_4 (u'+v')-i k^0_2 (v'+w')+i k^0_3 w'-i k^0_1 u'}\Big)  \nn \\
  &=\sum_{DDUU} 
     {\cal F} c_{++--}   \Big( \frac{-i}{(k^0_4 - k^0_1) (k^0_3 -k^0_2) (k^0_3 - k^0_1) } +\frac{-i}{(k^0_4 - k^0_1) (k^0_4 - k^0_2) (k^0_3 - k^0_2) } \Big)
    \Xi^{I}_{DDUU}  \nn \, ,
\end{align}
with $\Xi^{I}_{DDUU}  $ given in \cref{Xi} below.

In this way we can evaluate all terms in ${\cal J}^I_\textsc{nlo}$. The result is 
\begin{align}
 {\cal J}^I_\textsc{nlo} &=\sum_{DUUU}
            i {\cal F} \,\Xi ^I_{DUUU}  A_{DUUU} \(-c_{+---}  + c_{-+--}- c_{-++-}+c_{+-+-}  \)
                  \nn\\
& +\sum_{DDDU} 
 i {\cal F} \, \Xi ^I_{DDDU} A_{DDDU}
                          \(c_{--+-}- c_{+-+-} -c_{-++-}+c_{+++-}   \)
                          \nn\\
                          &+\sum_{DDUU}
         i  {\cal F}\, \Xi ^I_{DDUU} A_{DDUU}
                            \(-c_{-+--} +c_{++--}+   c_{--+-}- c_{+-+-}\) \, ,
\label{J1res}                            
\end{align}
with ${\cal F}$ defined in \cref{F} and
\begin{align}
  A_{DUUU} &= \frac{1}{(k^0_1 - k^0_2) (k^0_1 - k^0_3) (k^0_1 - k^0_4)}\, ,  \nn \\
  A_{DDDU} &=\frac{1}{(k^0_4 - k^0_1) (k^0_4 - k^0_2) (k^0_4 - k^0_3)}\, ,  \nn \\
A_{DDUU} &=\frac{1}{(k^0_1 - k^0_3) (k^0_1 - k^0_4) (k^0_2 -k^0_3)} +\frac{1}{(k^0_4 - k^0_1) (k^0_4 - k^0_2) (k^0_2 - k^0_3) }\, .
\end{align}
 For the CP conserving source $\Xi_i^\textsc{cp} =1$ are trivial, while for the CP violating source 
\be
\Xi^{\CPV}_{DUUU} =\frac{-i (2 k^0_1 - k^0_2 - k^0_4)}{(k^0_1 - k^0_2) (k^0_1 - k^0_4)},\quad
  \Xi^{\CPV}_{DDDU}= \frac{i(2k^0_4 -k^0_1 - k^0_3)}{(k^0_4 - k^0_1) (k^0_4-k^0_3)},\quad
  \Xi^{\CPV}_{DDUU} =\frac{ i(k^0_1 - k^0_2 + k^0_3 - k^0_4)}{(k^0_2 - k^0_3) (k^0_1 - k^0_4)}.
  \label{Xi}
  \ee
The $c$-factors in \cref{J1res} can be rewritten in terms of the distribution functions
\begin{align}
  \(-c_{+---}  + c_{-+--}- c_{-++-}+c_{+-+-}  \)&=-c_{+-}(k^0_1,k^0_2)=-(g^>_1 g^<_2 -g^<_1 g^>_2)\, , \nn \\
  \(c_{--+-}- c_{+-+-} -c_{-++-}+ c_{+++-}   \) &=\;\;\;c_{+-}(k^0_3,k^0_4)=\;\;\; (g^>_3 g^<_4 -g^<_3 g^>_4)\, , \nn \\
  \(-c_{-+--} +c_{++--}+   c_{--+-}- c_{+-+-}\)&=\;\;\;c_{+-}(k^0_3,k^0_2)=\;\;\;(g^>_3 g^<_2 -g^<_3 g^>_2)\, ,
 \label{cfactors}                                               
\end{align}
where we used that $(g^>_i-g^<_i)=1$. The $c_{+-}$-terms defined in \cref{c_pm} are expanded in small chemical potential as in \cref{c0}.

\subsection{Relaxation rate}

For the CP conserving source, the leading order term in the small $\mu$ expansion cancels, just as for the LO calculation. To see this, divide the summation in $(a_1,a_2,a_3,a_4) \in \overline{DUUU}$ and the opposite half $(\bar a_1,\bar a_2,\bar a_3,\bar a_4)\in \underline{DUUU}$, as discussed in \cref{s:contour}.  For zero chemical potential the  first term in \cref{J1res} is proportional to
\begin{align}
\Re \big[ i\sum_{DUUU}  {\cal F} A_{DUUU}c_{0,+-} \big]
  &=   \Re \big[ i \sum_{\overline{DUUU}} \big( 
                {\cal F} A_{DUUU}c_{0,+-} +\sum_{\overline{DUUU}}  
                {\cal F} A_{DUUU}c_{0,+-} \big)\big] \nn \\
  &=   \Re \big[ i\sum_{\overline{DUUU}}  \big(
                {\cal F} A_{DUUU}c_{0,+-} +
   ( {\cal F} A_{DUUU}c_{0,+-} \big)^*  \big) \big] =0\, ,
    \label{cancel_NLO}
\end{align}
where we used that for zero chemical potential $g^> (-k^0) = - g^<(k^0)$. Similarly, the other terms in ${\cal J}_\textsc{nlo}^\textsc{cp}$ vanish at leading order.  Hence, the first contribution comes from the term linear in the chemical potential.  We can again divide the  summation into two halves, and using that $h(-k^0) = h(k^0)$ now they add:
\begin{align}
\Re \big[ i\sum_{DUUU}  {\cal F} A_{DUUU}c_{1,+-}  \big]
  &=-2\Im \big[ \sum_{\overline{DUUU}}  i
    {\cal F} A_{DUUU}c_{1,+-} \big] \, .
    \label{add_NLO}
\end{align}

Putting it all together we find that
\begin{align}
   S_\textsc{nlo}^\textsc{cp}&=2s|f|^4 \Re \int_{ \vec k} \,  {\cal J}^\textsc{cp}_\textsc{nlo}
\nn\\
  &= \frac{4 |f|^4 }{T} \Im \int_{ \vec k} \, \Big[
    \sum_{\overline{DUUU}}
     {\cal F} A_{DUUU} (\mu_1 h(k_1) - \mu_2 h(k_2))
                      - \!\!\!\sum_{\overline{DDDU}}  
      {\cal F} A_{DDDU}  (\mu_3 h(k_3) - \mu_4 h(k_4))
\nn \\
 &\hspace{2.2cm}-\sum_{\overline{DDUU}} 
    {\cal F} A_{DDUU} 
   (\mu_3 h(k_3) - \mu_2 h(k_2)) \Big] \, ,
\end{align}
with $\mu_1 = \mu_3 =\mu_L$ and $\mu_2 = \mu_4 =\mu_R$.

We first do the sum over  the $k^0_i$ appearing in the $h$-functions.  For the first term, we choose the set $\overline{DUUU} =(1,1,a_3,a_4), \, (2,1,a_3,a_4) = (\E_L^*,\E_R,a_3,a_4) ,\,(-\E_L,\E_R,a_3,a_4)$ with $a_3,a_4$ unconstrained. The remaining sum is over the poles in $k_3^0 =k_L^0$ and $k_4^0 =k_R^0$.  Similarly, for the second term we choose   $\overline{DDDU}= (a_1,a_2,\E_L^*,\E_R) ,\,(a_1,a_2,-\E_L,\E_R)$, and for the third term  $\overline{DDUU}= (a_1,-\E_R,-\E_L^*,a_4) , \,(a_1,-\E_R,\E_L,a_4)$.

We write
\be
 \Gamma_\textsc{nlo}^\pm = \pm\frac{ |f|^2 }{2T} \Im  \int_{\vec k} \frac{1}{ \omega_L \omega_R }  \,
 \[\Delta^\textsc{cp}_1  \frac{  \(h(\E_L^*) \mp h(\E_R) \)}{\E_L^*  -\E_R} \tr_1^\textsc{lo}
   +  \Delta^\textsc{cp}_2   \frac{ \( h(\E_L) \mp  h(\E_R) \)}{\E_L+\E_R} \tr_2^\textsc{lo} \] \label{eq:relaxNLO} \, ,
  \ee
where we factored out the leading order trace factors $\tr_i^\textsc{lo}$ defined in \cref{tr0}.
The coefficients $\Delta^\textsc{cp}_i$ are
\begin{align}
   \frac{4\omega_L \omega_R}{|f|^2}   \Delta^\textsc{cp}_1 &=     \sum_{UU} 
    \frac{ (-1)^{\sum F_i} \tr_{1a}^\textsc{nlo} }{ (\E_L^* - k^0_L) (\E_L^* - k^0_R)}
 +\sum_{DD}  
          \frac{ (-1)^{\sum F_i}\tr_{1a}^\textsc{nlo} }{ (\E_R - k^0_L) (\E_R - k^0_R) }
          \nn \\
  &\hspace{2cm}+\sum_{DU} \frac{(-1)^{\sum F_i} \tr_{1b}^\textsc{nlo}}{(k_L^0 - k_R^0) }
    \Big(\frac{-1}{\E_L^* +k_L^0  }+\frac{1}{ k_R^0 +\E_R }\Big) \, ,
\nn \\
    \frac{4\omega_L \omega_R}{|f|^2}  \Delta_2^\textsc{cp} &=     \sum_{UU} 
    \frac{ (-1)^{\sum F_i}\tr_{2a}^\textsc{nlo}}{ (\E_L + k^0_L) (\E_L+ k^0_R)}
 +\sum_{DD}  
    \frac{ (-1)^{\sum F_i}\tr_{2a}^\textsc{nlo} }{ (\E_R - k^0_L) (\E_R - k^0_R)} 
\nn \\
  &\hspace{2cm}+
 \sum_{DU} 
    \frac{(-1)^{\sum F_i}\tr_{2b}^\textsc{nlo}}{ (k_L^0 - k_R^0) } 
    \Big(  \frac{1}{\E_L-k_L^0  } +\frac{1}{k_R^0 +\E_R }\Big)\, ,
    \label{Delta1}
\end{align}
with $\tr ^\textsc{nlo}_i =1$ for scalars, and for fermions
\begin{align}
    \tr_1^\textsc{lo}\tr^\textsc{nlo}_{1a}& = \tr^\textsc{nlo}|_{\{k_1^0= \E_L^*,k_2^0 = \E_R\}}  ,  & \tr_1 ^\textsc{lo}\tr^\textsc{nlo}_{1b}& = \tr^\textsc{nlo}|_{\{k_1^0= {-\E_L^*},k_2^0 = {-\E_R}\}}, \nn\\
 \tr_2 ^\textsc{lo}  \tr^\textsc{nlo}_{2a} &= \tr^\textsc{nlo}|_{\{k_1^0= -\E_L,k_2^0 = \E_R \}},&\tr_2 ^\textsc{lo}\tr^\textsc{nlo}_{2b} &= \tr^\textsc{nlo}|_{\{k_1^0= {\E_L},k_2^0 = {-\E_R}\}} \, .
\label{tr1a}                                                                                                                         
\end{align}    
with $\tr^\textsc{nlo}$ given in \cref{tr1}.

Finally we do the summation over the remaining poles in $k_{L,R}^0$.  We find $\Delta^\textsc{cp} \equiv \Delta^\textsc{cp}_1 =\Delta^\textsc{cp}_2$ for both scalars and fermions with
\begin{align}
  \Delta^\textsc{cp}
  &= -\frac{|f|^2}{4\omega_L\omega_R} \frac{ 4 \( \Im(\E_L)+\Im(\E_R)\)^2 \Re(\E_L) \Re(\E_R)}{|\E_L+\E_R|^2 |\E_L-\E_R^*|^2 \Im(\E_L) \Im(\E_R) } {\tilde\tr^\textsc{nlo}}
\nn\\
&=-| f|^2\frac{ (\Gamma_L+\Gamma_R)^2}{\Gamma_L \Gamma_R\((\Gamma_L+\Gamma_R)^2+(\omega_L-\omega_R)^2\)\((\Gamma_L+\Gamma_R)^2+(\omega_L+\omega_R)^2\)}\tilde \tr^\textsc{nlo} \, .
     \label{DCP}
\end{align}
For scalars $\tilde \tr^\textsc{nlo}=1$ while for fermions
\begin{align}
  \tilde \tr^\textsc{nlo} &
                   = -\frac12\(\vec k^2 -\frac{\Gamma_L(\omega_R^2 + \Gamma_R^2) + \Gamma_R(\omega_L^2 + \Gamma_L^2) }{\Gamma_L +\Gamma_R}\).
                   \label{trDCP}
\end{align}
%

\subsection{CP violating source}

Since there is an extra factor $\Xi^{\CPV}$ in ${\cal J}_\textsc{nlo}^{\CPV}$ in \cref{J1res}, the leading order term in the small chemical potential expansion contributes for the CP-violating source. Just as in the LO calculation we can effectively set $\mu_i =0$. The summation over poles can again be divided into opposite halves, which add.  The result is then
\begin{align}
  S_\textsc{nlo}^{\CPV}
           & =   4|f|^2 \delta \Im \int_{ \vec k} \frac1{2^4 \omega_L^2\omega_R^2} \bigg[
   \sum_{\overline{DUUU}}
   \frac{(-1)^{\sum F_i} \tr^\textsc{nlo} (n(k^0_1)-n(k^0_2))  (2 k^0_1 - k^0_2 - k^0_4)}{(k^0_1 - k^0_2)^2 (k^0_1 - k^0_3) (k^0_1 - k^0_4)^2} 
                  \nn\\
&+\sum_{\overline{DDDU}} 
\frac{(-1)^{\sum F_i} \tr^\textsc{nlo} (n(k^0_3)-n(k^0_4))(2k^0_4 -k^0_1 - k^0_3)}{(k^0_4 - k^0_1)^2 (k^0_4 - k^0_2) (k^0_4 - k^0_3)^2}
                          \nn\\
              & +\sum_{\overline{DDUU}}
                \frac{(-1)^{\sum F_i} \tr^\textsc{nlo} (n(k^0_3)-n(k^0_2)) (k^0_1 - k^0_2 + k^0_3 - k^0_4)}{(k^0_2 - k^0_3)^2 (k^0_1 - k^0_4)^2 } \Big(
                \frac{1}{k^0_1 - k^0_3 } -\frac{1}{ k^0_4 - k^0_2  }\Big)
                \bigg]    \, .
\end{align}
We sum over $k^0_i$ appearing
in the number densities $n$; we make he same choices as for the CP-conserving source in the previous subsection. We can then write
\begin{align}
S_\textsc{nlo}^{\CPV} 
           &= - \delta \, \int_{\vec k} \frac1{ \omega_L\omega_R} 
             \Im\[
              \Delta_1^{\CPV} \, \, \frac{n(\E_L^*) -n(\E_R)}{(\E_L^*-\E_R)^2}\tr^\textsc{lo} _1
            + \Delta_2^{\CPV} \, \,
             \frac{n(\E_L) + n(\E_R)+s}{(\E_L+\E_R)^2} \tr^\textsc{lo} _2
 \] \, ,
\end{align}
with
\begin{align}
  \frac{4 \omega_L\omega_R}{ |f|^2}  \Delta^{\CPV} _1 &=    \sum_{UU} 
    \frac{ (-1)^{\sum F_i} \tr_{1a}^\textsc{nlo}  (2\E_L^*-\E_R-k_R)}{ (\E_L^* - k^0_L) (\E_L^* - k^0_R)^2}
 +\sum_{DD}  
          \frac{ (-1)^{\sum F_i}\tr_{1a}^\textsc{nlo}  (2 \E_R-k_L-\E_L^*)}{ (\E_R - k^0_L)^2 (\E_R - k^0_R) }
          \nn \\
  &\hspace{1.5cm}+\sum_{DU} \frac{(-1)^{\sum F_i} \tr^\textsc{nlo}_{1b}(k_L+\E_R-\E_L^*-k_R)}{(k_L^0 - k_R^0)^2 }
    \Big(\frac{1}{\E_L^* +k_L^0  }-\frac{1}{ k_R^0 +\E_R }\Big) \, ,
    \nn \\
   \frac{4 \omega_L\omega_R}{ |f|^2}  \Delta^{\CPV} _2 &= \sum_{UU} 
    \frac{ (-1)^{\sum F_i}\tr_{2a}^\textsc{nlo} (2\E_L+\E_R+k^0_R)}{ (\E_L + k^0_L) (\E_L+ k^0_R)^2}
 +\sum_{DD}  
    \frac{ (-1)^{\sum F_i}\tr_{2a}^\textsc{nlo}  (2 \E_R-k_L^0+\E_L)}{ (\E_R - k^0_R)  (\E_R - k^0_L)^2} 
\nn \\
  &\hspace{1.5cm}
 +\sum_{DU} 
    \frac{(-1)^{\sum F_i}\tr_{2b}^\textsc{nlo}  (k_L+\E_R+\E_L-k_R)}{(k_L^0 - k_R^0) ^2} 
    \Big(  \frac{1}{\E_L-k_L^0 } +\frac{1}{k_R^0 +\E_R }\Big) .
    \label{B_CPV}
\end{align}

We can do the final summation over poles in $\Delta_i^{\CPV}$, but unlike the CP conserving case where some terms cancel and the final result is reasonably simple, the result cannot be written in a concise way.  What is more, we now find $ \Delta_1^{\CPV}\neq \Delta_2^{\CPV}$ and also $\Delta_i^{\CPV}$ is generically complex. The NLO contribution can thus no longer be written as a simple additive factor in the integrand.

\section{Results and discussion}
\label{s:results}

In this section we summarize the main results for the NLO calculation, and discuss the implications.

\subsection{Relaxation rate}

The full relaxation rate up to NLO is
\be
 \Gamma^\pm = \pm\frac{ |f|^2 }{2T} \Im  \int_{\vec k} \frac{1}{ \omega_L \omega_R }  \,
 \[  \frac{  \(h(\E_L^*) \mp h(\E_R) \)}{\E_L^*  -\E_R} \tr_1^\textsc{lo} 
   +   \frac{ \( h(\E_L) \mp  h(\E_R) \)}{\E_L+\E_R} \tr_2^\textsc{lo} \]\( 1+\Delta^\textsc{cp}\)\, ,
 \label{G_summary}
  \ee
 with the NLO contribution captured by $\Delta^\textsc{cp}$ given in \cref{DCP,trDCP}.  The rate $\Gamma^-$ is dominated by the first term in the square brackets, which has a resonance in small thermal widths for degenerate masses. In this limit the expression for the rate and the NLO correction simplifies considerably, and we can derive analytical expressions.
  
To see the resonance, we write $m_L^2 = m_R^2 +\delta m^2$ and expand in small mass difference $|\delta m^2| \ll m_L^2$. The first term between square brackets in the expression for $\Gamma^-$ in \cref{G_summary} becomes
  \be
  \Im \Big[\frac{h(\E_L^*) + h(\E_R)}{\E_L^*  -\E_R}\Big]
  = \frac{h(\omega_L)}{\Gamma_L\Big( 1 + \(\frac{\delta m^2}{4\Gamma_L \omega_L}\)^2 \Big)}+{\cal O}\left(\frac{\Gamma_L}{\omega_L}, \frac{\delta m^2}{\omega_L^2}\right) \, ,
  \label{expansion}
  \ee
where we additionally approximated $\Gamma_R \approx \Gamma_L$, and in the denominator assumed $\Gamma_L \ll T$ as is appropriate for a perturbatively generated thermal width. 
If in addition $|\delta m^2| \ll 4\Gamma_L T$, the mass difference in the denominator can be neglected. This is an excellent approximation for the quarks. On the other hand, for leptons the $\delta m^2$-term in the denominator in \cref{expansion} dominates.  Since $\delta m^2 \ll T^2$ this term is still enhanced compared to the 2nd term between square brackets in \cref{G_summary}, but the parametric dependence of the resonance is different.  The second term between brackets in $\Gamma^-$ and both terms in $\Gamma^+$ do not have this resonant structure and are subdominant. Indeed, expanding in small mass difference gives
\begin{align}
  \Im\Big[ \frac{h(\E_L^*) - h(\E_R) }{\E_L^*  -\E_R}\Big]={\cal O}\left(\frac{\Gamma_L}{\omega_L},\frac{\delta m^2}{\omega_L^2}\right),\qquad
  \Im\Big[ \frac{ h(\E_L) \mp  h(\E_R) }{\E_L+\E_R}\Big]=  {\cal O}\left(\frac{\Gamma_L}{\omega_L},\frac{\delta m^2}{\omega_L^2}\right)\, .
\end{align}

It thus follows that the rate  $\Gamma^-$ is resonantly enhanced in the degenerate mass limit. The NLO contribution $\Delta^\textsc{cp}$ is an additive factor to the integrand, and thus also to the resonant factor in the integrand; it does not affect or shift the resonance structure itself up to ${\cal O}(\frac{\Gamma_L}{\omega_L},\frac{\delta m^2}{\omega_L^2})$ corrections.

To estimate the NLO contribution to $\Gamma^-$  we will for simplicity consider the exactly degenerate case, a good approximation for quarks, and write
\be
m_T \equiv m_L = m_R,\quad \Gamma_T\equiv \Gamma_L= \Gamma_R\, , \quad \omega_T \equiv \omega_L = \omega_R\, ,
\label{degenerate}
\ee
with thermal width $\Gamma_T^2 \ll T^2$.
Then
\begin{align}
  \Delta^\textsc{cp}
  &= -\frac{|f|^2}{4\Gamma_T^2(\omega_T^2+\Gamma_T^2) }\tilde \tr^\textsc{nlo} \approx - \frac{|f|^2}{4\Gamma_T^2\omega_T^2}\,\tilde \tr^\textsc{nlo} \, .
\end{align}
The traces are trivial for scalars, and $\tr_1^\textsc{lo}=  2\,\tilde \tr^\textsc{nlo}               = (m_T^2 +\Gamma_T^2)$
for fermions. Putting it all together we get for scalars and fermions respectively
\begin{align}
  \Gamma_s^-
  &= \frac{|f_s|^2}{2T} \int_{\vec k} \frac{1 }{\omega_T^2}
 \Bigg[\frac{ h_s(\omega_T)}{\Gamma_T} + ...\Bigg]
\Big(1- \frac{|f_s|^2}{4\omega_T^2\Gamma_T^2} + ... \Big) \, ,
\nn \\
\Gamma_f^-
  &= \frac{|f_f|^2}{2T}\int_{\vec k} \frac{1 }{\omega_T^2}
 \Big[\frac{ h_f(\omega_T) (m_T^2+\Gamma_T^2)}{\Gamma_T} + ... \Big]
    \Big(1-\frac{|f_f|^2 (m_T^2+\Gamma_T^2)}{8\omega_T^2\Gamma_T^2} + ... \Big) \, ,
    \label{G_estimate}
\end{align}
where the ellipses denote  ${\cal O}(\frac{\Gamma_L}{\omega_L},\frac{\delta m^2}{\omega_L^2})$ corrections. 
When the NLO contribution dominates and $|\Delta_i^\textsc{cp}|>1$, the relaxation rate becomes negative, an unphysical result, signalling the breakdown of the vev-insertion approximation. This is expected to be cured when the full tower of the higher order terms in the coupling expansion is added, which amounts to a resummation of the spacetime-dependent mass.  

Whether the VIA expansion is valid depends on the details of the model. To get some insight we estimate the thermal mass and width 
as $m_T^2 \sim \alpha T^2$, $\Gamma_T \sim \alpha T$, with $\alpha = \frac{g^2}{4\pi}$ with $g$ the largest relevant (gauge) coupling. The integral in \cref{G_estimate} is dominated by momenta $\omega_T \sim T$. 

Let's first look at scalars. In supersymmetric set-ups,  the interaction with the Higgs is typically trilinear and we parametrize $|f_s| = \frac1{\sqrt{2}} A_s \vp_b$. The relaxation rate is maximized in the broken phase with $\vp_b=\vp_N$, the Higgs vev at nucleation. For a strong first order phase transition we need $\vp_N/T_N \gtrsim 1$, with $T_N$ the nucleation temperature. During the phase transition we then find
\be
|\Delta_s^\textsc{cp}| \sim \frac{ A_s^2\vp_N^2}{8 \alpha ^2 T_N^4} \sim \frac{1}{8\alpha^2} \frac{A_s^2}{T_N^2} .
\ee
For neutralinos the thermal corrections are set by the weak  gauge interactions and $\alpha \approx 0.03$. The NLO contribution is important unless the trilinear coupling is significantly smaller than the electroweak scale $A_s < {\cal O}(0.1) T_N$.  We conclude that VIA breaks down in these type of SUSY models.

For fermions with a Yukawa type Higgs interaction we parametrize $|f_f| \approx  \tilde y_f \vp_b$. Possible corrections from higher order Higgs interactions, although maybe essential for CP-violation, can be neglected in $|f_f|$.  If only the SM Higgs field obtains a vev during the phase transition then $\tilde y_f =y_f$ equals the SM Yukawa coupling.  We estimate
\be
|\Delta_f^\textsc{cp}| \approx \frac{|\tilde y_f|^2 \vp_N^2 }{8T_N^2\alpha }\sim \frac{|\tilde y_f|^2}{8\alpha} \, .\label{dEstimate}
\ee
For quarks with strong interactions $8\alpha \sim 1$ and we find that for $\tilde y_f \sim 1$ the NLO contribution becomes important.  If the phase transition is aligned with the SM Higgs direction and  $\tilde y_f =y_f$, the perturbative expansion breaks down for the top quark which has $y_{t} =1$; the calculation is well in the perturbative regime for all other quarks. For larger values of $\varphi_N/T_N$, though this is hard to achieve in actual phase transitions, the vev-insertion approximation breaks down for smaller values of $\tilde y_f$.

Going beyond the degenerate mass limit requires a numerical evaluation of \cref{G_summary,gamma0,eq:relaxNLO}.  For this we
choose the representative values $\vp_N =T_N =100\,$GeV.  For the thermal masses of the (up-type) quarks and leptons 
we use \cite{Chung:2009cb}\footnote{The mass of the doublet $m_{q,L} $ is taken as the average of the up- and down quark mass, and we have neglected the small contribution from the down quark.}
\begin{align}
  m_{q,L}^2 &= T^2 \bigg(\frac{g_1^2}{288} + \frac{3 g_2 ^2}{32} + \frac{g_3^2}{6} 
              + \frac{\tilde y_f^2}{16} \bigg)\, ,  \qquad
	&& m_{q,R}^2 =  T^2 \left(\frac{g_1^2}{18} + \frac{g_3^2}{6}  + \frac{\tilde y_f^2}{8} \right)\, , \\
	m_{l,L}^2 &= T^2 \left(\frac{g_1^2}{32} + \frac{3 g_2 ^2}{32} + \frac{\tilde y_f^2}{16} \right)\, ,  \qquad
	&& m_{l,R}^2=  T^2 \left(\frac{g_1^2}{8}  + \frac{\tilde y_f^2}{8} \right)\, ,
\end{align}
where $g_1,g_2$ and $g_3$ denote the $U(1)$, $SU(2)$ and $SU(3)$-gauge couplings respectively.
For the thermal widths, we use \cite{Joyce:1994zn}
\begin{equation}
  \Gamma_{q} = \frac 4 3 \alpha_s T \approx 0.16 T \, ,\qquad \Gamma_{l,L} = \alpha_w T \approx \frac{T}{30}\, , \qquad \Gamma_{l,R} = \frac 3 2 \alpha_w \tan^2 \theta_w T \approx \frac{T}{70}\,,
  \label{estimate2}
\end{equation}
for quarks, left-handed and right-handed leptons respectively.

\begin{figure}
\centering
	\includegraphics[width = 0.7\textwidth]{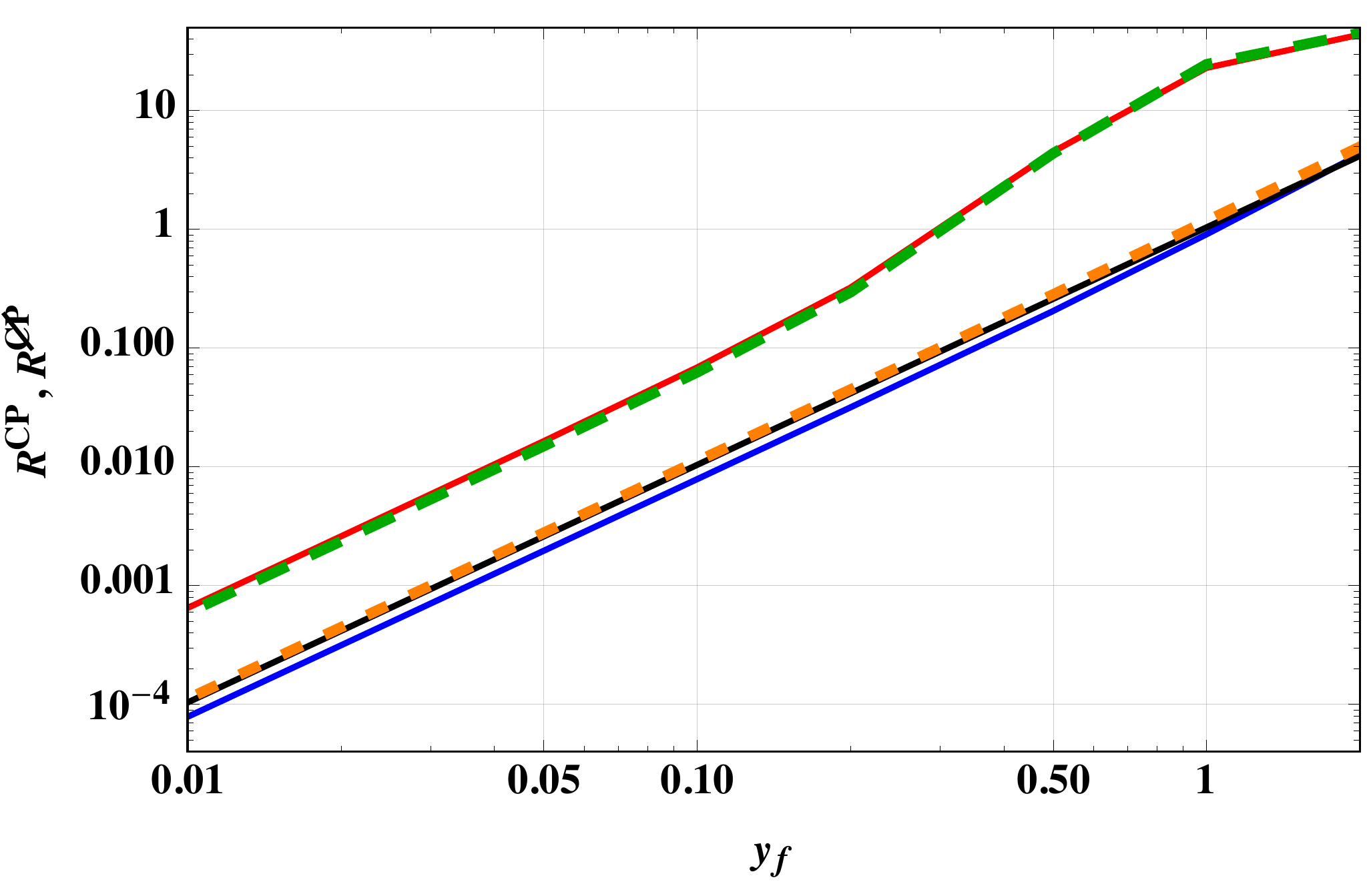}
	\caption{ Ratio of the NLO and LO contributions  to the relaxation rate $R^\textsc{cp} \equiv |\Gamma^-_\textsc{nlo}|/|\Gamma^-_\textsc{lo}|$ and source term $R^\textsc{\CPV} \equiv |S^{\CPV}_\textsc{nlo}|/|S^{\CPV}_\textsc{lo}|$, evaluated deep in the broken phase. The blue (red) line shows $R^\textsc{cp}$ for quarks (leptons), and the dashed orange and green lines show $R^{\CPV}$ for quarks and leptons respectively. The black line shows the estimate of $R^\textsc{cp}$ and $R^{\CPV}$ for quarks in the mass degenerate limit, corresponding to \cref{G_estimate} and \cref{S_estimate} with $\Delta^\textsc{cp}, \Delta^{\CPV}$ estimated by \cref{dEstimate}. We take $\varphi_N = 100 \, \text{GeV} $ and $T = 100 \, \text{GeV}$ . } \label{fig:ratios}
\end{figure}

In \cref{fig:ratios} we show the ratio of the NLO and LO contribution to the relaxation rate $R^\textsc{cp} \equiv |\Gamma^-_\textsc{nlo}|/|\Gamma^-_\textsc{lo}|$ deep into the broken phase (where $\varphi_b \rightarrow \varphi_N$) as a function of the Yukawa couplings for quarks and leptons in red and blue respectively.  The estimate of \cref{G_estimate}, with \cref{dEstimate} for the quarks in the mass degenerate limit is shown in black and matches the full result up to a factor $\mathcal O (1.5)$.  The good agreement between the numerical and analytical calculations confirms that the relaxation rate is dominated by the resonance.  We indeed find that the (absolute value of the) NLO result for quarks becomes comparable to the LO result for $\tilde y_f \gtrsim 1$.

For leptons the ratio $R^\textsc{cp}$ is approximately a factor 8 larger than for quarks for the same coupling, and as a result the VIA breaks down already at smaller coupling $\tilde y_f = {\cal O}(0.1)$. Non-perturbativity is never an issue for leptons if the bounce is along the SM direction, but may still be in other set-ups that have $\tilde y_f \gg y_f$.  As discussed above, for leptons the exactly degenerate limit \cref{degenerate} and thus the estimate \cref{G_estimate} is not a a good approximation, but we can nevertheless understand why $R^\textsc{cp}$ is larger. For quarks the NLO correction \cref{estimate2} is enhanced by $\alpha^{-1}$, with $\alpha$ the QCD coupling constant.  For leptons we similarly expect that the NLO correction scales with inverse powers of $\alpha$, now arising from inverse powers of $\delta m^2$ rather than the thermal width.  For leptons, though, the thermal corrections are set by the weak interactions and $\alpha$ is the weak coupling constant, giving a larger enhancement.

\subsection{CP violating source}

Adding the LO and NLO contribution to the CP-violating source term gives
\begin{align}
S^{\CPV} 
           &= - \delta \, \Im \int_{\vec k} \frac1{ \omega_L\omega_R} 
             \Im\[
             \frac{n(\E_L^*) -n(\E_R)}{(\E_L^*-\E_R)^2}\tr^\textsc{lo}_1 \, (1+\Delta_1^{\CPV}) 
       +
             \frac{n(\E_L) + n(\E_R)}{(\E_L+\E_R)^2} \tr^\textsc{lo}_2  \, (1    + \Delta_2^{\CPV}) 
 \] \, ,
\end{align}
where we removed the divergent term by normal ordering, and 
with the coefficients $\Delta_i^{\CPV}$ given in \cref{B_CPV,tr1a}.  Just as for the relaxation rate, the first term is resonantly enhanced in the degenerate mass limit
\be
\Im \Big[\frac{n(\E_L^*) -n(\E_R)}{(\E_L^*-\E_R)^2}\Big] = \frac{h(w_T)}{2 T\, \Gamma_T} + {\cal O}\left( \frac{\Gamma_T}{\omega_T} \right),\quad
\Im \Big[\frac{n(\E_L) + n(\E_R)}{(\E_L+\E_R)^2}\Big] = {\cal O}\left(\frac{\Gamma_T}{\omega_T}\right) \, .
\ee

Unlike the relaxation rate, $\Delta_1^{\CPV} \neq \Delta_2^{\CPV}$ and $\Im[\Delta_i^{\CPV}] \neq 0$, and the NLO contribution is in general not simply an additive part in the integrand.  However, the resonant term only depends on $\Delta_1^{\CPV} $ which  is a real factor in the mass degenerate limit \cref{degenerate}.  In particular, we then find
\begin{align}
S_s^{\CPV} 
           &= - \frac{\delta_s}{2T} \, \int_{\vec k} \frac1{ \omega_T^2} 
       \Big (     \frac{h_s(w_T)}{ \Gamma_T } + ...\Big)
             \, \Big (1+\frac{ |f_s|^2 }{4 \omega_T^2\Gamma_T^2 }  +
... \Big) 
            \, , \nn \\ 
  S_f^{\CPV} 
           &= - \frac{\delta_f}{2T}  \, \int_{\vec k} \frac1{ \omega_T^2} 
            \Big (\frac{h_f(w_T)m_T^2}{\Gamma_T} +... \Big)
             \, \Big (1+\frac{ |f_f|^2 m_T^2}{8 \omega_T^2\Gamma_T^2} +
          ...\Big)  \, , \label{S_estimate}
\end{align}
where the ellipses denote  ${\cal O}(\frac{\Gamma_L}{\omega_L},\frac{\delta m^2}{\omega_L^2})$ corrections.
Apart from the overall sign the corrections are exactly the same as for the relaxation rate \cref{G_estimate}. For fermions the VIA approximation thus also breaks down for the calculation of the CP-odd source for  couplings $\tilde y = {\cal O}(1)$ for quarks.

\begin{figure}
\centering
	\includegraphics[width = 0.7\textwidth]{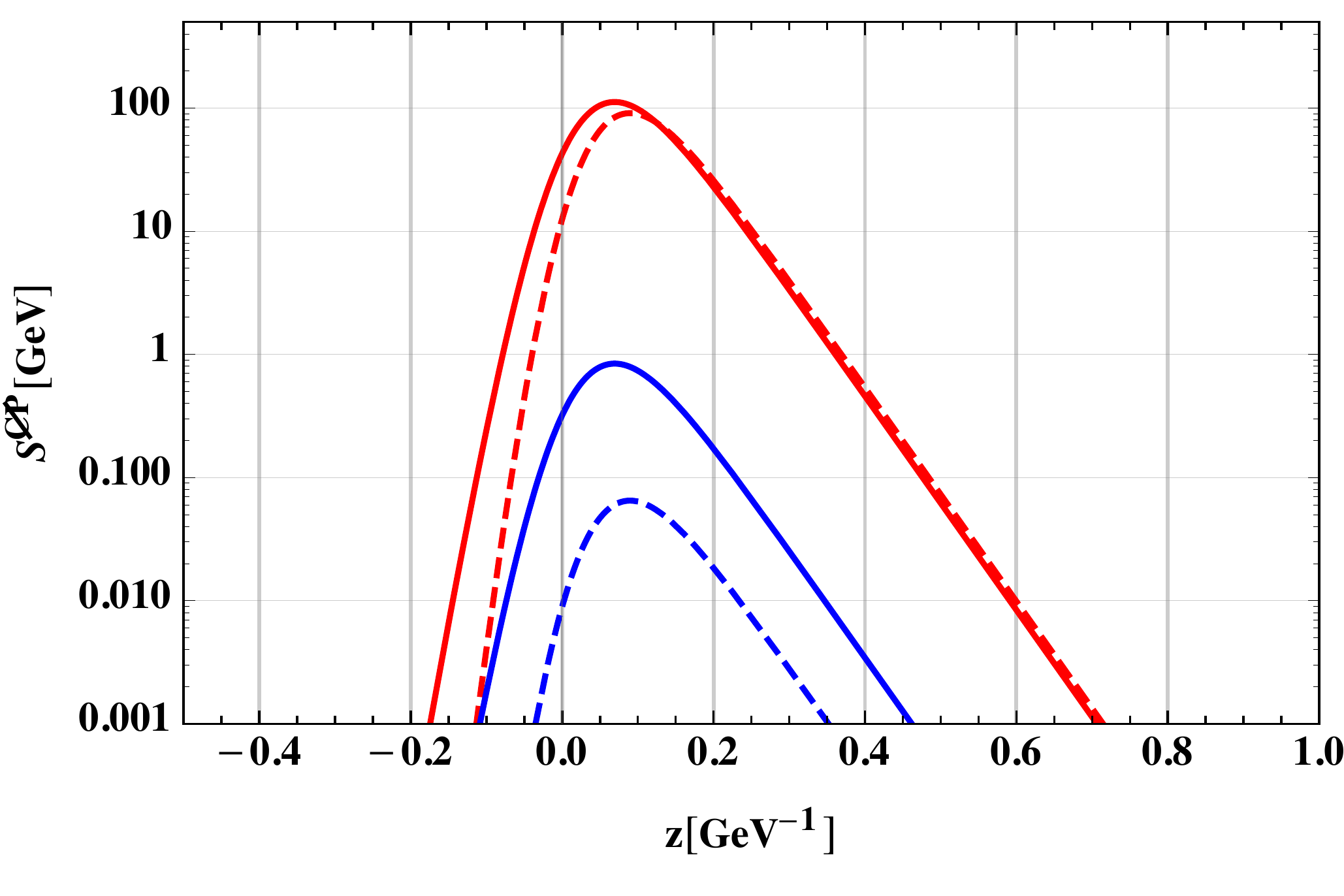}
	\caption{$z$-dependence of the LO (solid lines) and NLO (dashed lines) source terms for a quark with $\tilde y_f = 0.1$ (red) and $\tilde y_f = 1$ (blue). We use the bubble wall profile  $\vp_b(z)= \varphi_N/2 (1 + \tanh{z/L_w} )$, with $L_w = 10/T$\, $\varphi_N = T_N = 100\, \text{GeV}$ and $|f_f| \approx \tilde y_f \varphi_b$. CP-violation stems from a dimension-6 operator as in Refs. \cite{Zhang:1994fb, Bodeker:2004ws, Balazs:2016yvi, deVries:2017ncy}, giving $\delta = 3 \tilde y_f^2 v_w \varphi_b^3 \varphi_b' /\Lambda^2$, where the prime denotes a derivative with respect to $z$, we take $\Lambda = 1\, \text{TeV}$ and bubble wall velocity $v_w = 0.05$.}	\label{fig:sourcebroken}
\end{figure}

\Cref{fig:ratios,fig:sourcebroken} show the numerical results for the source. In \cref{fig:ratios} we plot the ratio $R^\textsc{\CPV} \equiv |S^\textsc{\CPV}_\textsc{NLO}|/|S^\textsc{\CPV}_\textsc{LO}|$ for quarks (dashed orange) and leptons (dashed green) as a function of the coupling $\tilde y_f$. Not surprisingly giving our estimate in the degenerate limit, the ratio is nearly identical to $R^\textsc{CP}$ for the relaxation rates.  Although the source terms is space-time dependent and varies over the bubble wall, this cancels out in the ratio $R^\textsc{\CPV}$ which is evaluated deep in the broken phase.
In \cref{fig:sourcebroken} we plot the LO and NLO contribution across the bubble wall for a quark for two different choices of the Yukawa coupling.   For this plot we took the bubble profile $\vp_b(z)= \frac{\varphi_N}{2} (1 + \tanh{z/L_w} )$. The CP-violation stems from a complex dimension-6 Yukawa-like interaction and $f_f = \tilde y_f \vp_b(1+ i \vp^2/\Lambda^2)$ with $\Lambda $ the cutoff scale.  Evaluating the $\Delta^{\CPV}$ with $\vp_b =\vp_N$ gives a good estimate for the size of the NLO correction in the $z$-region where the source peaks.

\subsection{Discussion}

To summarize, we have shown that the NLO contribution to the relaxation rates and source terms is small as long as $|f_s|/T_N^2 \lesssim  \alpha$ for scalars and $|f_f|/T_N \lesssim \sqrt{\alpha}$ for fermions, with $\alpha$ the QCD (electroweak) coupling for scalars/fermions with strong (only electroweak) interactions. 
For larger effective couplings $|f_i|$ the vev-insertion approximation breaks down.

Focussing on specific implementations, in a supersymmetric setup with CP violation in the neutralino sector, this implies that VIA breaks down for trilinear couplings $|f_s| \approx A_s \vp_b$ for $A_s > {\cal O}(0.1)T_N$. For fermions with a Yukawa type couplings $|f_f| \approx \tilde y_f \vp_b$ the breakdown happens for  $\tilde y_f > {\cal O}(1)$ for quarks, and   $\tilde y_f > {\cal O}(0.1)$ for leptons. This means in particular that if only the SM Higgs field obtains a vev during the phase transition, the source terms cannot be reliably computed for the top quark, but a baryogenesis scenario with CP violation in the lepton sector  \cite{Joyce:1994bi, Joyce:1994zn, Chung:2009cb, Chiang:2016vgf, Guo:2016ixx, deVries:2018tgs} is under calculational control.

We finally note that the resonant behavior of the relaxation rate and source term is preserved at next-to-leading order.

\section*{Acknowledgements}
The authors thank T. Konstandin for useful discussions and J. de Vries for valuable comments on the manuscript. MP and JvdV have received support from the Netherlands Organization for Scientific Research (NWO). JvdV is funded by the Deutsche Forschungsgemeinschaft under Germany`s Excellence Strategy -- EXC 2121 ``Quantum Universe'' -- 390833306.

\appendix
\newpage

\renewcommand{\theequation}{\thesection.\arabic{equation}}
\numberwithin{equation}{section}


\bibliographystyle{h-physrev3} 
\bibliography{myrefs}

\end{document}